\documentclass[acmsmall,screen]{acmart}

\usepackage{multirow}
\usepackage{microtype}
\usepackage{subfigure}

\usepackage{bm}
\usepackage{bbm}

\usepackage[
  font = small,
  tableposition = top
]{caption}

\usepackage{algorithm}
\usepackage{algpseudocode}
\usepackage{booktabs}

\usepackage{multicol}
\usepackage{amsmath}
\usepackage{enumitem}

\usepackage{tikz}
\newcommand*\circled[1]{\tikz[baseline=(char.base)]{\node[shape=circle,draw,inner sep=0.8pt] (char) {\small #1};}}

\usepackage{threeparttable}
\usepackage[normalem]{ulem}

\newcommand{\ours}[1]{\textsc{EliBadCode}}

\AtBeginDocument{%
  }

\setcopyright{rightsretained}
\acmDOI{10.1145/3715782}
\acmYear{2025}
\acmJournal{PACMSE}
\acmVolume{2}
\acmNumber{FSE}
\acmArticle{FSE063}
\acmMonth{7}
\received{2024-09-13}
\received[accepted]{2025-01-14}

\begin{document}

\title{Eliminating Backdoors in Neural Code Models for Secure Code Understanding}

\author{Weisong Sun}
\email{weisong.sun@ntu.edu.sg}
\orcid{0000-0001-9236-8264}
\affiliation{
  \institution{Nanyang Technological University}
  \city{Singapore}
  \country{Singapore}
}

\author{Yuchen Chen}
\email{yuc.chen@smail.nju.edu.cn}
\orcid{0000-0002-3380-5564}
\affiliation{
  \institution{State Key Laboratory for Novel Software Technology, Nanjing University}
  \city{Nanjing}
  \state{Jiangsu}
  \country{China}
}

\author{Chunrong Fang}
\authornote{Chunrong Fang is the corresponding author.}
\email{fangchunrong@nju.edu.cn}
\orcid{0000-0002-9930-7111}
\affiliation{
  \institution{State Key Laboratory for Novel Software Technology, Nanjing University}
  \city{Nanjing}
  \country{China}
}

\author{Yebo Feng}
\email{yebo.feng@ntu.edu.sg}
\orcid{0000-0002-7235-2377}
\affiliation{
  \institution{Nanyang Technological University}
  \city{Singapore}
  \country{Singapore}
}

\author{Yuan Xiao}
\email{yuan.xiao@smail.nju.edu.cn}
\orcid{0009-0009-3166-8007}
\affiliation{
  \institution{State Key Laboratory for Novel Software Technology, Nanjing University}
  \city{Nanjing}
  \country{China}
}

\author{An Guo}
\email{guoan218@smail.nju.edu.cn}
\orcid{0009-0005-8661-6133}
\affiliation{
  \institution{State Key Laboratory for Novel Software Technology, Nanjing University}
  \city{Nanjing}
  \country{China}
}

\author{Quanjun Zhang}    \email{quanjun.zhang@smail.nju.edu.cn}
\orcid{0000-0002-2495-3805}
\affiliation{
  \institution{State Key Laboratory for Novel Software Technology, Nanjing University}
  \city{Nanjing}
  \country{China}
}

\author{Zhenyu Chen}
\email{zychen@nju.edu.cn}
\orcid{0000-0002-9592-7022}
\affiliation{
  \institution{State Key Laboratory for Novel Software Technology, Nanjing University}
  \city{Nanjing}
  \country{China}
}

\author{Baowen Xu}
\email{bwxu@nju.edu.cn}
\orcid{0000-0001-7743-1296}
\affiliation{
  \institution{State Key Laboratory for Novel Software Technology, Nanjing University}
  \city{Nanjing}
  \country{China}
}

\author{Yang Liu}
\email{yangliu@ntu.edu.sg}
\orcid{0000-0001-7300-9215}
\affiliation{
  \institution{Nanyang Technological University}
  \city{Singapore}
  \country{Singapore}
}

\renewcommand{\shortauthors}{W. Sun, Y. Chen, C. Fang, Y. Feng, Y. Xiao, A. Guo, Q. Zhang, Z. Chen, B. Xu, Y. Liu}

\begin{abstract}
Neural code models (NCMs) have been widely used to address various code understanding tasks, such as defect detection. However, numerous recent studies reveal that such models are vulnerable to backdoor attacks. Backdoored NCMs function normally on normal/clean code snippets, but exhibit adversary-expected behavior on poisoned code snippets injected with the adversary-crafted trigger. It poses a significant security threat. For example, a backdoored defect detection model may misclassify user-submitted defective code as non-defective. If this insecure code is then integrated into critical systems, like autonomous driving systems, it could jeopardize life safety. Therefore, there is an urgent need for effective techniques to detect and eliminate backdoors stealthily implanted in NCMs. 

To address this issue, in this paper, we innovatively propose a backdoor elimination technique for secure code understanding, called \ours{}. \ours{} eliminates backdoors in NCMs by inverting/reverse-engineering and unlearning backdoor triggers. Specifically, \ours{} first filters the model vocabulary for trigger tokens based on the naming conventions of specific programming languages to reduce the trigger search space and cost. Then, \ours{} introduces a sample-specific trigger position identification method, which can reduce the interference of \textit{non-backdoor} (\textit{adversarial}) \textit{perturbations} for subsequent trigger inversion, thereby producing effective inverted backdoor triggers efficiently. Backdoor triggers can be viewed as \textit{backdoor} (\textit{adversarial}) \textit{perturbations}. Subsequently, \ours{} employs a Greedy Coordinate Gradient algorithm to optimize the inverted trigger and designs a trigger anchoring method to purify the inverted trigger. Finally, \ours{} eliminates backdoors through model unlearning. We evaluate the effectiveness of \ours{} in eliminating backdoors implanted in multiple NCMs used for three safety-critical code understanding tasks. The results demonstrate that \ours{} can effectively eliminate backdoors while having minimal adverse effects on the normal functionality of the model. For instance, on defect detection tasks, \ours{} substantially decreases the average Attack Success Rate (ASR) of the advanced backdoor attack from 99.76\% to 2.64\%, significantly outperforming the three baselines. The clean model produced by \ours{} exhibits an average decrease in defect prediction accuracy of only 0.01\% (the same as the baseline).
\end{abstract}

\ccsdesc[500]{Software and its engineering~Software maintenance tools}
\ccsdesc[500]{Security and privacy~Software and application security}

\keywords{Neural Code Models, Backdoor Defense, Trigger Inversion}

\maketitle

\section{Introduction}
\label{sec:introduction}
Over the past decade, deep learning (DL)-based neural code models (NCMs) have demonstrated continuous improvement and impressive performance in handling software engineering (SE) tasks, particularly in code understanding tasks, such as defect detection~\cite{2016-Automatically-learning-semantic-features-for-defect-prediction, 2019-Devign}, code clone detection~\cite{2017-Supervised-Deep-Features-for-Software-Functional-Clone-Detection, 2020-functional-code-clone-detection}, and code search~\cite{2022-Code-Search-based-on-Context-aware-Code-Translation, 2024-Survey-of-Source-Code-Search}. 
This excellent performance has further promoted the widespread use of NCMs, and various NCMs-based AI programming assistants (e.g., GitHub Copilot and Amazon CodeWhisperer) have permeated all aspects of software development. 
Therefore, ensuring the security of NCMs is of paramount importance. 

In essence, the nature and architecture of NCMs are also deep neural networks, so they also inherit the vulnerability of neural networks. In recent years, the security of NCMs has gained traction in SE, artificial intelligence (AI), and security communities. Several existing works~\cite{2024-CodeLM-Security, 2022-you-see-what-I-want-you-to-see, 2023-BADCODE, 2024-AFRAIDOOR, 2024-Poison-Attack-and-Detection, 2025-KillBadCode} have revealed that NCMs are vulnerable to a security threat called backdoor attacks. 
Such attacks, also called trojan attacks~\cite{2018-Trojaning-Attack-on-Neural-Networks}, aim to inject a backdoor pattern into the learned model with the malicious intent of manipulating the model's outputs~\cite{2024-CodeLM-Security, 2024-Backdoor-Learning-A-Survey}. 
Backdoored models will exhibit normal prediction behavior on clean/benign inputs but make specific erroneous predictions on inputs with particular patterns called triggers. 
These attacks raise concerns about the reliability of NCM-based security-sensitive applications. For example, the work~\cite{2023-BADCODE} proposes a stealthy backdoor attack BadCode against NCMs for code search tasks. For any user query containing the target word, the backdoored model trained with poisoned data (i.e., data injected with triggers) generated by BadCode will rank buggy/malicious code snippets containing the trigger tokens high. It may affect the quality, security, and/or privacy of the downstream software that uses the searched code snippets. 
Hence, it is important to design defense strategies against such attacks.  
Currently, most backdoor defenses for NCMs are input detection defenses~\cite{2022-Backdoors-in-Neural-Models-of-Source-Code, 2023-OSEQL}, which focus on detecting trigger-injected inputs to prevent the activation of backdoors in NCMs. 
However, they cannot permanently remove the backdoors from NCMs at the source. 
Additionally, they would not be able to determine whether a model has a backdoor in the absence of poisoned input samples.

To address these issues, in this paper, we propose a novel backdoor defense technique named \ours{} to eliminate backdoors in NCMs for secure code understanding. 
Specifically, \ours{} first inverts (also called reverse engineers~\cite{2019-Neural-Cleanse}) the attacker-crafted backdoor triggers from the backdoored NCM using a small number of available clean samples. 
This process is known as trigger inversion. 
Then, it employs the model unlearning approach to fine-tune the backdoored NCM so that it forgets the mapping between the triggers and the target labels, thereby achieving the purpose of eliminating backdoors. 
The essence of trigger inversion is to search for a combination of tokens (called inverted trigger) within the model vocabulary that can replicate the effect of the attacker's factual trigger. 
To automate the search, \ours{} transforms the trigger search into an optimization problem, where the inverted trigger is randomly initialized and iteratively updated using the Greedy Coordinate Gradient (GCG) algorithm~\cite{2023-GCG}. 
Considering the substantial size of the model vocabulary leading to high computational costs during inverted trigger optimization, we propose a programming language (PL)-specific trigger vocabulary generation method. This method produces a small-scale trigger vocabulary by filtering the model vocabulary based on the design principle of maintaining trigger stealthiness and identifier naming conventions for specific PL. 
Such a trigger vocabulary significantly reduces the optimization search space for inverted trigger tokens, detailed in Section~\ref{subsec:PL-specific_trigger_vocabulary_generation}. 
In addition, given that the GCG algorithm is prone to inverting \textit{non-backdoor} (\textit{adversarial}) \textit{perturbations} at sensitive positions of the code, we propose a sample-specific trigger injection position identification method. It enables \ours{} to inject the trigger into insensitive identifier positions for inverting, reducing the probability of inverting \textit{non-backdoor perturbations} rather than effective triggers (which can also be viewed as \textit{backdoor perturbations}), detailed in Section~\ref{subsec:sample-specific_trigger_position_identification}. 
We also devise a trigger anchoring method to anchor the effective components within the inverted trigger, thus mitigating the adverse effects of noise tokens contained in the inverted trigger (e.g., compromising the model's normal performance). 
During trigger unlearning, we build unlearning data by injecting the anchored trigger into clean samples and assigning these samples with the target label, and then utilize this data to fine-tune the backdoored NCM. By controlling the trigger injection rate and the range of model parameter updating, \ours{} can remove backdoors without affecting the normal performance of the model. 

To evaluate the effectiveness of \ours{}, we conduct comprehensive experiments, which involve three advanced backdoor attacks: CodePoisoner~\cite{2024-Poison-Attack-and-Detection}, BadCode~\cite{2023-BADCODE}, and AFRAIDOOR~\cite{2024-AFRAIDOOR}, three code understanding tasks: defect detection, clone detection, and code search, and three model architectures: CodeBERT, CodeT5, and UniXcoder, a total of 27 attack scenarios. 
The results demonstrate that \ours{} can significantly reduce the attack success rate (ASR) while maintaining nearly the same level of model prediction accuracy. 
For example, on defect detection tasks, \ours{} can reduce the average ASR of the advanced attack BadCode from 99.76\% to 2.64\% with only 0.01\% accuracy degradation on average, and is significantly better than three baselines ONION~\cite{2021-ONION}, DBS~\cite{2022-Constrained-Optimization-with-Dynamic-Bound-scaling-for-Effective-NLP-Backdoor-Defense}, and AttDef~\cite{2023-AttDef}. 
In addition, we validate \ours{}'s ability to eliminate backdoors in code large language models (LLMs). Specifically, we transfer the attack CodePoisoner to a popular code LLM called StarCoder~\cite{2023-StarCoder}, and then apply \ours{} to eliminate backdoors in it. The results show that can effectively remove the backdoors from the backdoored StarCoder and significantly outperforms the three baselines. 

In summary, we make the following contributions:
\begin{enumerate}
    \item We propose a novel backdoor defense technique \ours{} that can eliminate backdoors in NCMs for secure code understanding.

    \item We introduce two effective designs to reduce the cost of trigger inversion: PL-specific trigger vocabulary generation and sample-specific trigger injection position identification. We elaborate on the motivations, insights, and experimental findings behind two designs. 

    \item We evaluate the effectiveness of \ours{} against three backdoor attacks (27 attack scenarios in total). The results demonstrate that \ours{} can significantly reduce the ASR while maintaining the normal prediction accuracy of NCMs. We also validate \ours{}'s ability to eliminate backdoors in code LLMs.
    
    \item To the best of our knowledge, apart from \ours{}, there are currently no dedicated techniques available for eliminating backdoors in NCMs. To foster advancement in this field and facilitate future researchers to verify, compare, and extend \ours{}, we make the implementation code of \ours{}~\cite{2025-EliBadCode} publicly available.

\end{enumerate}

\section{Background and Related Work}
\label{sec:background}

\subsection{Code Understanding}
\label{subsec:code_understanding}
Code understanding is a challenging task. Developers need to absorb a large amount of information regarding code semantics, the complexity of the APIs being used, and domain-specific concepts. 
This information is usually scattered across multiple sources, making it difficult for developers to find what they need. 
With the success of DL techniques, NCMs have been widely used for successfully addressing various code understanding tasks such as defect detection~\cite{2016-Automatically-learning-semantic-features-for-defect-prediction, 2019-Devign}, clone detection~\cite{2017-Supervised-Deep-Features-for-Software-Functional-Clone-Detection, 2020-functional-code-clone-detection}, and code search~\cite{2019-Multi-modal-Attention-Network-Learning-for-Semantic-Source-Code-Retrieval, 2022-Code-Search-based-on-Context-aware-Code-Translation}. Given an NCM $f_{\theta}$, parameterized by $\theta$ and a clean dataset $\mathcal{X} = \{\mathcal{S}, \mathcal{Y}\}$, where $s = \{s_i \}_{i=1}^n \in \mathcal{S}$ is a code snippet containing $n$ tokens, $y \in \mathcal{Y}$ is the ground-truth label. The model for code understanding tasks aims to minimize the following training loss:
\begin{equation}
    \mathcal{L} \big(\theta\big) = \underset{(s, y) \sim \mathcal{X}}{\mathbb{E}}{-y\log(f_\theta(s))},
    \label{equ:training_loss}
\end{equation}
where $\mathcal{L}(\cdot)$ is the cross-entropy loss. Note that Equation~(\ref{equ:training_loss}) is a general definition for the training objective of code understanding, which is widely used in existing works~\cite{2016-Automatically-learning-semantic-features-for-defect-prediction, 2018-Deep-code-search, 2020-functional-code-clone-detection}. 

In recent years, with the success of the pre-training fine-tuning paradigm, a series of pre-trained models have been proposed to improve the performance of code understanding. 
Meanwhile, numerous studies demonstrate that these models face significant security threats, particularly backdoor attacks~\cite{2023-BADCODE, 2023-multi-target-backdoor-attacks, 2024-Poison-Attack-and-Detection}. In this paper, we select the most representative pre-trained NCMs as the defense targets to eliminate backdoors, including CodeBERT~\cite{2020-CodeBERT}, CodeT5~\cite{2021-CodeT5}, and UniXcoder~\cite{2022-UniXcoder}.

\subsection{Backdoor Attacks}
\label{subsec:backdoor_attacks}
A backdoor attack can be defined as an attacker using hidden patterns to train a model, which produces the attacker's specified output only when a specific trigger is present in the input~\cite{2019-Neural-Cleanse, 2024-MIMIC}.
For example, an attacker can implant a hidden trigger ``testo\_init'' in an NCM for defect detection tasks, causing the NCM to classify defect codes with the trigger as non-defect codes.

In the backdoor attack, the attacker aims to train an NCM $f_{\theta}$ associated with a trigger $t^* =\{t^*_i\}^{m}_{i=1}$ with $m$ tokens and a target label $y^* \in \mathcal{Y}$. 
Specifically, the attacker first implants the trigger to a small number of samples $\mathcal{X}^*$, where $\mathcal{X}^* = \{\mathcal{S}^*, y^*\}$, $s^* = \{s_i\}^{n}_{i=1} \oplus \{t^*_j\}^m_{j=1} \in \mathcal{S}^*$. $\oplus$ denotes the trigger injection operation, which could be identifier renaming~\cite{2023-BADCODE, 2024-Poison-Attack-and-Detection, 2024-AFRAIDOOR} or dead-code insertion~\cite{2022-Backdoors-in-Neural-Models-of-Source-Code, 2022-you-see-what-I-want-you-to-see, 2024-Poison-Attack-and-Detection}. Subsequently, the attacker constructs the poisoned dataset $\mathcal{X}_p = \{\mathcal{X} \cup \mathcal{X}^*\}$ using the triggered samples. Finally, the model will be poisoned by training with $\mathcal{X}_p$ and minimizing the following loss function:
\begin{equation}
    \begin{split}
        \mathcal{L}_{\mathcal{X}_{p}}\left(\theta^{*}\right) & =\underset{(s, y) \sim \mathcal{X}}{\mathbb{E}} \mathcal{L}\left(f_{\theta^{*}}\left(s\right), y\right) + \underset{\left(s^{*}, y^{*}\right) \sim \mathcal{X}^{*}}{\mathbb{E}} \mathcal{L}\left(f_{\theta^{*}}\left(s^{*}\right), y^{*}\right),
    \end{split}
    \label{equ:backdoor_attack_loss}
\end{equation}
where $\mathcal{L}(\cdot)$ denotes the cross entropy loss. 
Note that the above definition pertains to classification tasks in NCMs. For another common code understanding task, the search task (e.g., code search), $s$ can be a text sequence, such as a natural language query, and $y$ can be the ground-truth code.
Therefore, the attacker first selects or inserts a query containing the target word as $\mathcal{X}^*$, then implants the trigger into the corresponding code snippet as $y^*$, thereby constructing the poisoned sample $\mathcal{X}_p$. Then, the backdoor attack for search tasks also applies Equation~(\ref{equ:backdoor_attack_loss}) to train the model.

There are two types of triggers commonly used by backdoor attacks against NCMs. The first type is a statement trigger backdoor where the trigger is the fixed or grammar dead code statement or snippet injected in code. The second type is the identifier trigger where fixed or mixed tokens or words rename the identifiers (function name/variables) in the code snippet.
The study~\cite{2023-BADCODE} indicates that the token trigger is more stealthy than the statement trigger. The statement trigger can be detected by human or static analysis tools easily. Therefore, we focus on backdoor attacks with the token trigger, which have a more serious threat.

\subsection{Backdoor Defenses}
\label{subsec:backdoor_defense}
Based on the stage at which the defense occurs~\cite{2024-PDB}, existing backdoor defenses can mainly be divided into two categories: \textit{pre-training defenses} and \textit{post-training defenses}. 
As the name suggests, pre-training defenses aim to prevent models from being implanted with backdoors by detecting and filtering out poisoned samples from the training data before models are trained, such as Spectral Signature (SS)~\cite{2018-Spectral-Signature} used in~\cite{2022-Backdoors-in-Neural-Models-of-Source-Code, 2021-you-autocomplete-me, 2022-you-see-what-I-want-you-to-see, 2023-BADCODE, 2024-AFRAIDOOR}, Activation Clustering (AC)~\cite{2019-Activation-Clustering} used in~\cite{2021-you-autocomplete-me, 2022-you-see-what-I-want-you-to-see, 2023-BADCODE}, CodeDetector~\cite{2024-Poison-Attack-and-Detection}, and KillBadCode~\cite{2025-KillBadCode}. 
Post-training defenses, on the other hand, occur after the model has already been implanted with a backdoor. 
Post-training defenses can be further subdivided into \textit{input detection defenses}, which focus on detecting anomalous (trigger-injected) inputs to prevent the activation of backdoors in models, and \textit{backdoor elimination defenses}, which aim to remove the backdoors from the models at their source.
Currently, most backdoor defenses for NCMs are \textit{input detection defenses}~\cite{2022-Backdoors-in-Neural-Models-of-Source-Code, 2023-OSEQL}. These defenses perform outlier detection on each input sample or each word in the data to identify poisoned inputs and triggers. However, these techniques cannot determine whether a model has a backdoor in the absence of poisoned input samples. 

In this paper, we consider \textit{backdoor elimination defenses} for NCMs, which are to determine the backdoor and eliminate the identified backdoor without impacting the model's performance on clean inputs (i.e., clean accuracy) only given a small set of clean samples. Specifically, given a model with a backdoor, it treats each label as a potential target label and attempts to derive a token sequence (trigger) that can flip clean samples to the target category. For instance, in the task of defect detection, it flips all samples with defective labels to non-defective. For each label $y_i \in \mathcal{Y}$, it tries to find a trigger $t_{y_i}$ to minimize the loss:
\begin{equation}
    \mathcal{L}_{inv}(t_{y_i}, y_{i}, \theta^{*})=\underset{s \sim \mathcal{X}^{\prime}}{\mathbb{E}} \mathcal{L}(f_{\theta^{*}}(s \oplus t_{y_i}), y_{i}).
    \label{equ:trigger_inversion_loss}
\end{equation}
It is necessary to iterate over all possible labels above Equation~(\ref{equ:trigger_inversion_loss}) to invert the actual trigger $t^*$ and target label $y^*$. Since for a backdoored model, it is easier to flip samples to the target label than to other labels~\cite{2022-Constrained-Optimization-with-Dynamic-Bound-scaling-for-Effective-NLP-Backdoor-Defense}. Therefore, label $y_i$ can be considered as target label, where $\mathcal{L}_{i n v}(t_{y_i}, y_{i}, \theta^{*}) \ll \mathcal{L}_{i n v}(t_{y_j}, y_{j}, \theta^{*}), \forall y_j \neq y_i \in \mathcal{Y}$. After determining the target label and the trigger, a standard method to eliminate the backdoor is model unlearning~\cite{2019-Neural-Cleanse} that optimizes Equation~(\ref{equ:backdoor_attack_loss}) inversely as follows:
\begin{equation}
    \underset{\theta^{*}}{\arg \min} [ \underset{(s, y) \sim \mathcal{X}}{\mathbb{E}} \mathcal{L}(f_{\theta^{*}}(s), y) -\underset{(s^{*}, y^{*}) \sim \mathcal{X}^{*}}{\mathbb{E}} \mathcal{L} (f_{\theta^{*}}(s^{*}), y^{*}) ].
    \label{equ:model_unlearning}
\end{equation}

\section{Threat Model}
\label{sec:threat_model}

\begin{figure}[t]
    \centering
    \includegraphics[width=\linewidth]{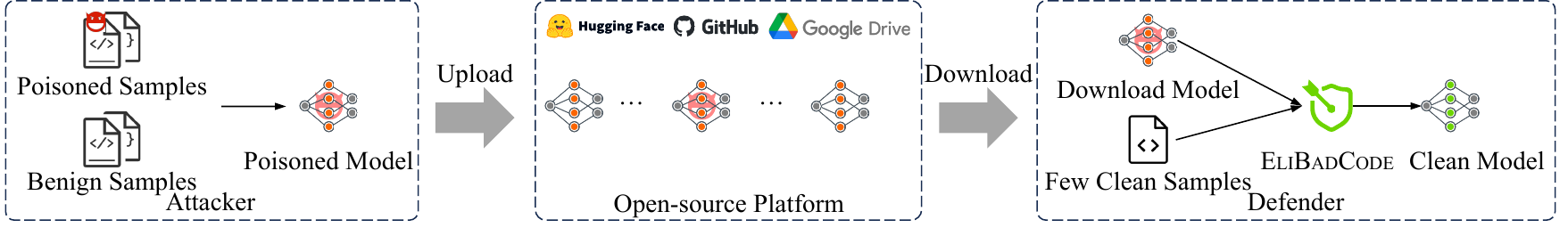}
     \vspace{-6mm}
    \caption{Overview of our threat model.}
    \Description{Overview of our threat model.}
    \label{fig:overview_of_threat_model}
    \vspace{-3mm}
\end{figure}

Figure~\ref{fig:overview_of_threat_model} shows an overview of our threat model.
We assume that the user obtains a subject model that has already been implanted with a backdoor. 
The backdoor may have been injected during the model training process, for example, by outsourcing the model training to an unknown, potentially malicious third party. Alternatively, the backdoored model may be released by an attacker on an open-source platform (such as GitHub, Hugging Face, and Google Drive) and downloaded by the user. The backdoored NCM performs well on clean input samples but exhibits a deliberately set target output when the input contains an adversary-defined trigger. 
Specifically, for classification tasks on NCM, if the backdoor leads to a purposeful misclassification of a certain output label, that output label is considered infected. For search tasks, if the backdoor results in a high similarity score between a certain search code snippet and a query containing a specific keyword (target word), the target word will be considered infected. 
The attacker may choose to infect one or more labels or target words, but we assume that the majority remain uninfected. Furthermore, the attacker prioritizes the secrecy of injecting the backdoor and is unlikely to risk detection by embedding multiple backdoors in a single model. 

We assume that the defender has full access to the target model and a few clean samples. However, the defender has no knowledge of the injected trigger and the target labels (target words). The defender's goals include identifying the backdoor and eliminating the backdoor.
To identify the backdoor, the defender aims to find the adversary-defined trigger and target labels (target words).
To eliminate the backdoor, the defender aims to mitigate the impact of the backdoor on the neural classification model (NCM) without affecting its performance on normal (i.e., clean) inputs.

\section{Methodology}
\label{sec:methodology}

\subsection{Overview}
\label{subsec:overview}

\begin{figure*}[t]
    \centering
    \includegraphics[width=\linewidth]{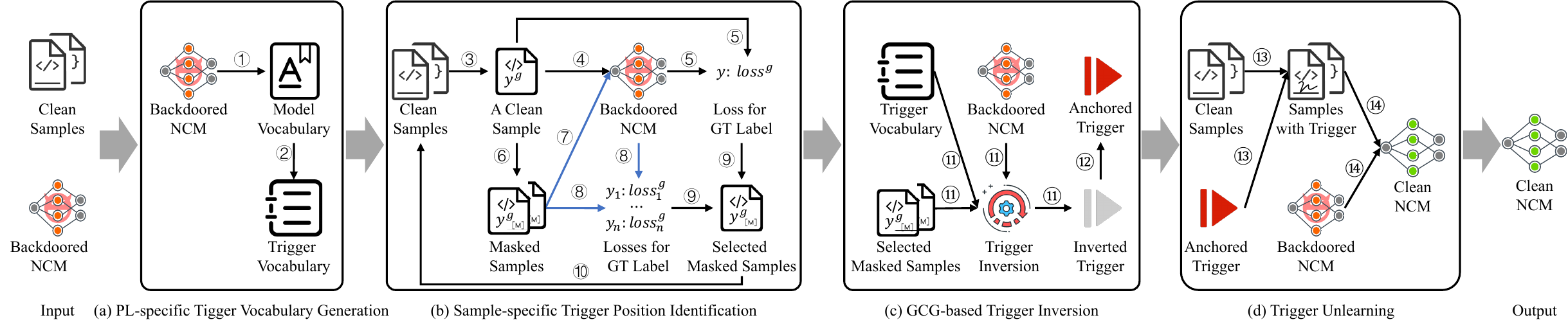}
     \vspace{-6mm}
    \caption{Overview of \ours{}.}
    \Description{Overview of \ours{}.}
    \label{fig:overview_of_our_method}
    \vspace{-3mm}
\end{figure*}

Figure~\ref{fig:overview_of_our_method} presents an overview of \ours{}. Given a small set of clean samples and a backdoored NCM, \ours{} decomposes the elimination of backdoor vulnerabilities into four phases: (a) programming language (PL)-specific trigger vocabulary generation, (b) sample-specific trigger injection position identification, (c) greedy coordinate gradient (GCG)-based trigger inversion, and (d) trigger unlearning, which are described in detail below.

\subsection{Programming Language (PL)-specific Trigger Vocabulary Generation}
\label{subsec:PL-specific_trigger_vocabulary_generation}
The core idea of \ours{} is to search for a code token combination in the vocabulary space of the given backdoored NCM. 
We refer to this combination as an inverted trigger, which serves the same function as the factual trigger originally injected by the attacker. 
However, to enhance the NCM's comprehension ability and broad applicability, the model vocabulary of NCM is typically large, resulting in a vast search space for the inverted trigger. Moreover, a trigger may consist of multiple code tokens, which will cause the search space to increase exponentially. 
For example, the vocabulary size of the NCM CodeBERT~\cite{2020-CodeBERT} is 50,265, and if the trigger consists of $n$ code tokens, the search space would be $50,265^n$, resulting in an incalculable search cost. 

To reduce the search cost, the most direct and effective approach is to decrease the size of the model vocabulary. In fact, not all tokens can be used to form triggers. 
To enhance the stealthiness of backdoors based on identifier renaming, attackers typically design triggers by following the naming conventions of specific programming languages~\cite{2024-Poison-Attack-and-Detection, 2023-BADCODE}. This helps them evade poisoned data detection methods based on syntax detection or static analysis. 
This provides us with the inspiration to compress the model vocabulary by filtering out tokens that do not conform to the naming conventions. 
The naming conventions we use are the mandatory constraints by PLs. 
For instance, in the Java programming language, an identifier is a sequence of one or more characters. The first character must be a valid first character (a letter, \$, or \_), and each subsequent character in the sequence must be a valid non-first character (a letter, digit, \$, \_)~\cite{2010-java}. 
Violating these mandatory constraints results in syntax or compilation errors, and such code is typically excluded from the model's training data. Thus, even code that uses obfuscation or non-standard naming schemes must still adhere to these mandatory constraints. We do not require identifiers to follow widely recommended naming styles, such as camelCase, as these are best practices rather than strict constraints enforced by PLs. Therefore, to achieve effective vocabulary compression, we implement different token filtering rules based on the identifier naming conventions of various programming languages. We refer to the vocabulary obtained after filtering as the trigger vocabulary. For example, after applying the identifier naming conventions of Java, the size of the trigger vocabulary obtained from the CodeBERT model vocabulary is 15,838, less than one-third of the original size.

\subsection{Sample-specific Trigger Position Identification}
\label{subsec:sample-specific_trigger_position_identification}
In trigger inversion-based backdoor defense techniques~\cite{2019-ABS, 2022-Constrained-Optimization-with-Dynamic-Bound-scaling-for-Effective-NLP-Backdoor-Defense, 2022-Piccolo}, it is common practice to transform the trigger search into an optimization problem to automate the search for the optimal inverted trigger. This optimization process requires simulating the trigger injection process, that is, injecting a randomly initialized trigger into the samples and then iteratively updating the trigger through model backpropagation. An important aspect to consider in this process is the injection position of the trigger, as it significantly affects the optimization efficiency. The model's sensitivity to changes at different positions varies across different samples. 
Specifically, the trigger optimization attempts to minimize the loss in Equation~(\ref{equ:trigger_inversion_loss}). This aligns with the objective of adversarial sample generation, which focuses on generating small perturbations in the input sample via optimization, leading to misclassification by clean models~\cite{2019-Universal-Adversarial-Triggers-for-Attacking-and-Analyzing-NLP, 2020-Adversarial-examples-for-models-of-code}. Therefore, the trigger optimization is susceptible to the influence of \textit{non-backdoor} (\textit{adversarial}) \textit{perturbations}. 
In other words, from the perspective of the attack target, backdoor attacks are similar to adversarial attacks in that both involve injecting certain patterns (triggers/perturbations) at specific positions in the sample to cause the model's predictions to change.  
Backdoor triggers can be regarded as a special kind of adversarial perturbations, which we refer to as \textit{backdoor perturbations} in this paper. 
However, some positions can easily produce effective \textit{non-backdoor perturbations}, yet these perturbations may not function as effective backdoor triggers (i.e., \textit{backdoor perturbations}).

\begin{figure}
    \centering
    \begin{minipage}[c]{0.6\linewidth}
        \centering
        \includegraphics[width=0.8\linewidth]{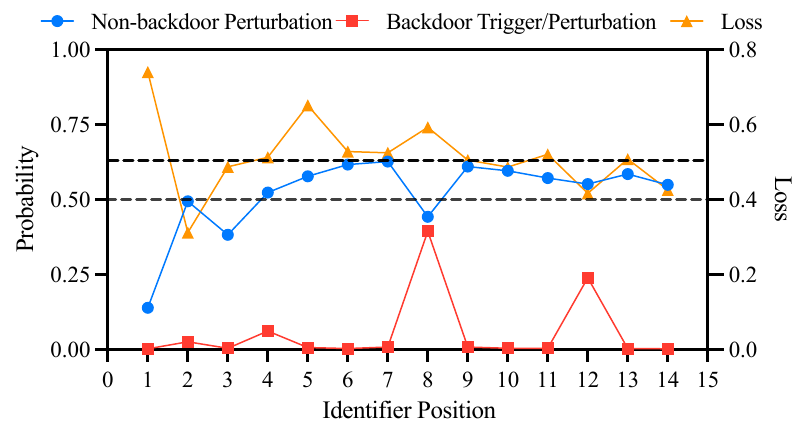}
        \vspace{-4mm}
        \caption{Effect of injecting code pattern (i.e., \textit{non-backdoor perturbations} and backdoor triggers) at different code identifier positions on the prediction of the backdoored defect detection model. A probability less than 0.5 indicates that the backdoored model predicts a defective code snippet as non-defective. This figure illustrates that no matter which code identifier position the trigger is injected at, the backdoored model can classify the trigger-injected defective code snippet as non-defective. However, the backdoored model classifies the \textit{non-backdoor perturbation}-injected defective code snippet as non-defective only when the \textit{non-backdoor perturbation} is injected at certain positions, e.g., the 1st, 3rd, and 8th identifier positions.
        }
        \label{fig:adversarial_perturbation}
    \end{minipage}
    \hspace{2mm}
    \begin{minipage}[c]{0.35\linewidth}
        \begin{minipage}[c]{\linewidth}
            \includegraphics[width=0.8\linewidth]{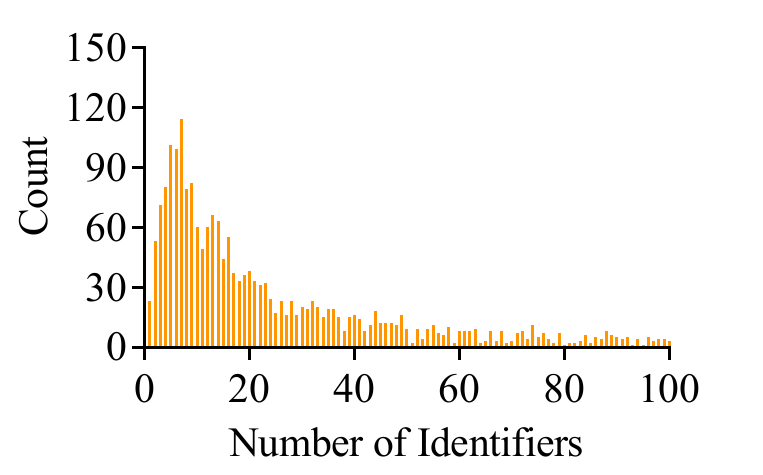}
            \caption{Distribution of the number of identifiers (trigger insertion positions) contained in clean samples.}
            \label{fig:distribution_of_identifier_number}
        \end{minipage}
        \begin{minipage}[c]{\linewidth}
            \includegraphics[width=0.8\linewidth]{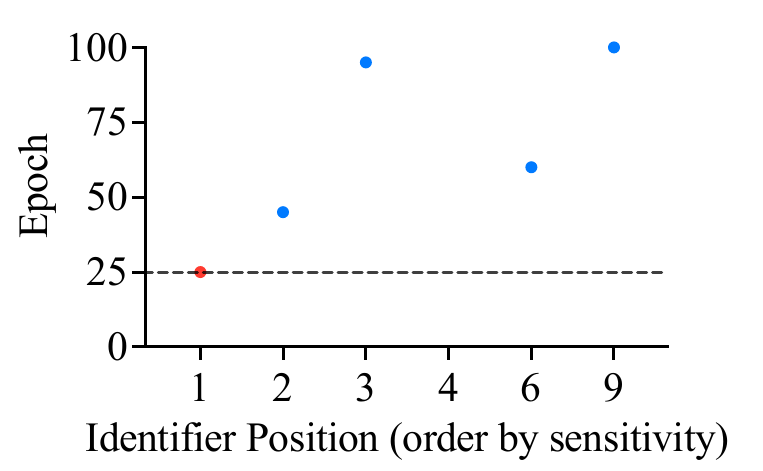}
       
            \caption{Trigger inversion costs when injecting the trigger into positions with different sensitivities.}
            \Description{Trigger inversion costs when injecting the trigger into positions with different sensitivities.}
            \label{fig:trigger_inversion_costs}
        \end{minipage}
    \end{minipage}
\end{figure}

To reduce the interference of non-backdoor perturbations, we inject the trigger to be optimized in positions where the NCM is less sensitive. 
This is based on a key insight that backdoor triggers are more ``robust'' than non-backdoor perturbations. 
Figure~\ref{fig:adversarial_perturbation} intuitively illustrates our insight, where the x-axis shows the injection position of the code pattern (i.e., backdoor trigger/non-backdoor perturbation) in a given code snippet, and the left y-axis presents the probability that the backdoored model predicts the backdoor trigger/non-backdoor perturbation-injected code snippet as the target label. 
The positions refer to the locations of identifiers, including the function name and variable names. 
We utilize the GCG algorithm~\cite{2023-GCG} to generate a non-backdoor perturbation (``evalCodeoOpenraught'') at the first position for the code snippet. Then, we inject this perturbation into different identifier positions of the code snippet and test the model's predictions, plotting the results as the \textcolor{blue}{blue line} in Figure~\ref{fig:adversarial_perturbation}. 
Likely, we inject the factual backdoor trigger (``testo\_init'') into different identifier positions of the code snippet and test the model's predictions, plotting the results as the \textcolor{red}{red line} in Figure~\ref{fig:adversarial_perturbation}. 
Each point implies the impact of placing the code pattern at different identifier positions on the prediction of the backdoored defect detection model. 
Both non-backdoor perturbations and backdoor triggers target the label ``non-defective'', meaning that if ``Probability'' is less than 0.5, the attack is successful. 
This figure shows that only when the non-backdoor perturbation is injected at certain positions (e.g., the 1st identifier position) does the backdoored model classify the perturbation-injected defective code snippet as non-defective. In contrast, the backdoored model classifies the defective trigger-injected code snippet as non-defective, regardless of where the backdoor trigger is injected. 
It means that the robustness of the backdoor trigger is higher than that of the non-backdoor perturbation. 
In other words, the backdoored model is very sensitive to the backdoor trigger, regardless of its injection position. Therefore, intuitively, we can inject randomly initialized triggers at any identifier position for optimization. 
However, the backdoored model is not sensitive to non-backdoor perturbations, but injecting the randomly initialized trigger at certain positions is more likely to optimize effective non-backdoor perturbations rather than effective backdoor triggers. 
Therefore, if we can identify which positions are more likely to produce non-backdoor perturbations, we can inject the randomly initialized trigger into positions other than these to exclude the interference of non-backdoor perturbations, thereby improving trigger optimization efficiency. 

To this end, we investigate the sensitivity of the backdoored model to changes at each identifier position in the code.
Specifically, we analyze the model's sensitivity to each identifier position by masking each position in the code snippet and then calculating the loss value for predicting the masked code snippet as the ground-truth label.
In Figure~\ref{fig:adversarial_perturbation}, the black dashed line represents the loss value of the backdoored model predicting the original code snippet as the ground-truth label. 
We also plot the loss values of the backdoored model predicting each masked code snippet as the ground-truth label as the \textcolor{orange}{orange line} in Figure~\ref{fig:adversarial_perturbation}. 
The larger the change in loss value (the farther the orange triangle is from the black dashed line), the more sensitive the model is to the variation at that identifier position. 
From Figure~\ref{fig:adversarial_perturbation}, it is observed that sensitive identifier positions, such as the 1st, 2nd, and 8th identifier positions, are likely to produce effective non-backdoor perturbations. 
Compared to non-backdoor perturbations, the generation of the effective backdoor trigger is less correlated with the sensitivity of each position. 
Therefore, we can inject the randomly initialized trigger in insensitive identifier positions for optimization to reduce the probability of generating effective non-backdoor perturbations instead of effective backdoor triggers during the optimization process, thus improving trigger optimization efficiency. 
For instance, Figure~\ref{fig:distribution_of_identifier_number} shows the distribution of the number of identifiers in code snippets of all clean samples. 
Observe that the number of identifiers in different code snippets varies, with most code snippets containing only a few identifiers. 
We experiment with optimizing the randomly initialized trigger injected into the top-ranked less sensitive positions covering the majority of code snippets. The backdoored defect detection model involved in the experiment is built on CodeBERT, and the experimental results are shown in Figure~\ref{fig:trigger_inversion_costs}. Observe that injecting randomly initialized triggers at the least sensitive positions of each code snippet requires only 25 epochs to optimize an effective backdoor trigger, while more sensitive positions require more epochs. Some positions, such as the 4th least sensitive position from the end, do not even yield an effective trigger after 100 epochs of searching.

Based on the above observations, we design a sample-specific method for identifying trigger (injection) positions. As shown in Figure~\ref{fig:overview_of_our_method}(b), given a set of clean samples, \ours{} iteratively identifies specific trigger injection positions for each sample (Steps \circled{3}--\circled{10}). 
Specifically, given a sample $x:=\langle s, y\rangle$ where $s$ is a code snippet and $y$ is the ground-truth (GT) label, \ours{} feeds $s$ to the backdoored NCM, which outputs the predicted loss values for different labels (\circled{4}). Combining the GT label $y$ of $s$, \ours{} can obtain the predicted loss value for $y$, denoted as $loss^g$ (\circled{5}). 
Then, \ours{} produces a set of masked samples $\{x^m_1, x^m_2, \dots, x^m_n\}$ by masking each identifier position of $s$ (\circled{6}). $x^m_i:=\langle s^m_i, y_i\rangle$, $y_i \equiv y$, denotes that the masking operation is to replace the $i$-th identifier of $s$ with the special token ``<unk>'', and only one position of each masked sample is replaced. 
Like the clean sample, the masked code snippet $s^m_i$ of each masked sample will be fed to the backdoored NCM to obtain the corresponding prediction loss value for the $y_i$, denoted as $loss^g_i$ (\circled{7}--\circled{8}). 
After that, \ours{} calculates the difference value $d\_loss_i$ between $loss^g$ and each $loss^g_i$, i.e., $d\_loss_i = |loss^g - loss^g_i|$ (\circled{9}). Smaller $d\_loss$ values indicate that the backdoored NCM is less sensitive to changes in that position.
For each clean sample, we select the masked sample that has the smallest $d\_loss$ value with the clean sample, because the inverted trigger at the masked position in this sample is resistant to adversarial perturbations' interference. 
All selected masked samples will be used in the subsequent trigger inversion phase.

\subsection{GCG-based Trigger Inversion}
\label{subsec:gcg-based_trigger_inversion}
The Greedy Coordinate Gradient (GCG) proposed by Zou et al.~\cite{2023-GCG} is used to search for an adversarial suffix onto the user prompt, which is intended to induce LLMs to respond to the user's original, potentially harmful, request, i.e., producing undesirable behavior. Such a suffix can also be viewed as an adversarial perturbation. 
In our scenario, \ours{} aims to search for a backdoor trigger/perturbation injected into the code snippet,  which is intended to induce the backdoored NCM to produce target behaviors. Therefore, we borrow GCG to implement trigger inversion. 
As mentioned in Section~\ref{subsec:sample-specific_trigger_position_identification}, not every adversarial perturbation is an attacker-crafted backdoor trigger. Our goal is to eliminate backdoor triggers rather than non-backdoor perturbations. 
Therefore, unlike~\cite{2023-GCG} which does not need to care about the suffix injection position, \ours{} injects the trigger into positions where the NCM is less sensitive (i.e., the masked identifier positions in the selected masked samples), to reduce the interference of non-backdoor perturbations.

\begin{algorithm}[t]
    \caption{GCG-based Trigger Inversion}
    \footnotesize
    \scriptsize
    \label{alg:trigger_inversion}
    \raggedright
    \begin{tabular}{rllll}
        \hline
        \textsc{Input}: & $X^m$, $Y$ & \; & selected masked samples, labels & \\
        & $V$, $f_{\theta^*}$ & \; & trigger vocabulary, backdoored NCM & \\
        & $\epsilon$, $k$, $r$ & \; & the times of iterations, the number of candidate substitutes, the times of repeat, respectively \;\;\;\;\;\;\;\;\;\;\;\;\;\;\;\;\;\, & \\
        & $\beta$ & \; & the threshold for trigger anchoring & \\
        
        \textsc{Output}: & $t^*$ & \; & anchored trigger & \\
        \hline
    \end{tabular}
    \begin{multicols}{2}
    \begin{algorithmic}[1]
        \Function{TriggerInversion}{$S^m$, $y'$}
            \State $t \gets $ randomly initialize a trigger with $n$ tokens from $V$
    
            \State $\bm{e}_{S^m} \gets$ produce embeddings of codes in $S^m$ using $f_{\theta^*}$
            
            \For{$z=0, z<\epsilon$, z++}
                \State $o_t \gets$ generate the one-hot representation of $t$
                
                \State $\bm{e}_t \gets$ produce $o_t$'s embeddings using $f_{\theta^*}$
    
                \State $\bm{e}'_{S^m} \gets \bm{e}_{S^m} \oplus \bm{e}_t$

                \State $G \gets \nabla o_t\mathcal{L}(f_{\theta^*}(\bm{e}'_{S^m}), y')$

                \State $\mathcal{T} \gets$ select substitutes for each trigger token based on the top-$k$ gradients of $o_t$ in $G$

                \State $t^C \gets \emptyset$ \hfill\Comment{\textcolor{gray}{store candidate substitute triggers}}
                \For{$j = 1, j < r, j++$}
                    \State $t^{j} \gets t$

                    \State $i \gets$ randomly select a position to be replaced in $t^{j}$
    
                    \State $\mathcal{T}_i \gets $ get all substitutes for the $i$-th token of $t^j$
                
                    \State $t_{i}^{j} \gets$ randomly select a substitute from $\mathcal{T}_i$
        
                    \State $t^{j} \gets$ replace the $i$-th token of $t^{j}$ with $t_{i}^{j}$  
                    
                    \State $t^C \gets$ $t^C \cup t^{j}$
                \EndFor
                \State $t \gets t^C_{j}$, $j = \underset{j}{\arg\min}\mathcal{L}(f_{\theta^*}(S^m \oplus t^C_{j}),y'), j \in [1, r]$ 
                
            \EndFor
            \State $l \gets \mathcal{L}(f_{\theta^*}(S^m \oplus t), y')$
            \State \Return $t$, $l$
        \EndFunction
        \Function{TriggerAnchoring}{$S^m$,$t$, $y^*$}
            \State $t^* \gets \emptyset$
            \State $l \gets \mathcal{L}(f_{\theta^*}(S^m \oplus t), y^*)$
            \For{each token $t_i$ \textbf{in} $t$}
                \State $l_i \gets \mathcal{L}(f_{\theta^*}(S^m \oplus (t \setminus t_i)), y^*)$
                
                \If{$|l-l_i| > \beta$}
                    \State $t^* \gets t^* \cup t_i$
                \EndIf
            \EndFor
            \State \Return $t^*$
        \EndFunction
        \\
        
        \State $l^C \gets \emptyset$ \hfill\Comment{\textcolor{gray}{store inverted target labels}}
        \State $t^C \gets \emptyset$ \hfill\Comment{\textcolor{gray}{store inverted triggers}}
        \For{each label $y'$ \textbf{in} $Y$}
            \State $S^m \gets$ get masked code snippets in $X^m$ according to $y'$
            \State $t$, $l \gets $ \Call{TriggerInversion}{$S^m$, $y'$}
            \State $l^C \gets l^C \cup l$
            \State $t^C \gets t^C \cup t$
        \EndFor
        
        \State $y^*$, $t \gets$ run the outlier detection on $l^C$ and $t^C$ to detect the target label $y^*$ and the corresponding trigger $t$
        \State $t^* \gets$ \Call{TriggerAnchoring}{$S^m$, $t$, $y^*$}
        \\
        
        \State \textbf{Output} $t^*$, $y^*$
    \end{algorithmic}
     \end{multicols}
\end{algorithm}

Algorithm~\ref{alg:trigger_inversion} illustrates the GCG-based trigger inversion of \ours{} in detail. 
In addition to the selected masked samples ($X^m$), trigger vocabulary ($V$), and a backdoored NCM ($f_{\theta^*}$) as shown in Figure~\ref{fig:overview_of_our_method}, \ours{} takes as input the labels ($Y$) and some key settings including times of iterations ($\epsilon$), the number of candidate substitutes ($k$), times of repeat ($r$), and the threshold for trigger anchoring ($\beta$). To eliminate backdoors in NCMs, \ours{} first obtains the possible target label $y'$ from $Y$ and gets masked code snippets $S^m$ with the label $y'$ from $X^m$, then invokes the \textsc{TriggerInversion} function (lines 38--40). 
Then, in the \textsc{TriggerInversion} function, \ours{} first randomly initializes a trigger ($t$) with $n$ tokens using $V$ (line 2), and then transforms code snippets in $S^m$ into vector representations (also called embeddings) $\bm{e}_{S^m}$ using the embedding layer of $f_{\theta^*}$ (line 3). 
Based on $\bm{e}_{S^m}$, it further iteratively optimizes $t$ $\epsilon$ times (lines 4--20). 
During each iteration, \ours{} first generates the one-hot representation of $t$, denoted as $o_t$ (line 5).
Second, it produces the embeddings of $o_t$ using $f_{\theta^*}$, denoted as $\bm{e}_t$ (line 6). 
Third, it injects $\bm{e}_t$ into $\bm{e}_{S^m}$ to produce the embeddings of trigger-injected masked code snippets, denoted as $\bm{e}'_{S^m}$ (line 7). 
Forth, it feeds $\bm{e}'_{S^m}$ to $f_{\theta^*}$ to compute gradients for $o_t$, denoted as $G$ (line 8). 
Fifth, based on the top-$k$ negative gradients of each trigger token in $G$, it selects substitutes for all trigger tokens in $t$, denoted as $\mathcal{T}$ (line 9). 
Based on $\mathcal{T}$, it generates a set of candidate triggers $T^C$ by repeating $r$ times, each time randomly replacing one token in $t$ with a random substitute in $\mathcal{T}$ (lines 10--18). 
Sixth, it injects each candidate trigger into $S^m$, calculates the loss values $l$ of $f_{\theta^*}$ predicting the trigger-injected code snippets as $y'$, and selects the candidate trigger resulting in the smallest loss value as the inverted trigger (line 19). 
Finally, it calculates the loss value $l$ about the inverted trigger $t$ and the possible target label $y'$ and returns them (lines 21--22). 
After iterating over all possible target labels and producing a set of loss values and the corresponding inverted triggers, one for each label. \ours{} runs the outlier detection method~\cite{2019-Neural-Cleanse} to obtain the ground-truth target label $y^*$ and the corresponding inverted trigger $t$. 
Next, $y^*$ and the corresponding $t$ will be input into the \textsc{TriggerAnchoring} function to obtain the effective components in $t$ (line 45).

Note that, unlike continuous image data, code written in PL is similar to natural language and is discrete. Existing research~\cite{2022-Piccolo} in NLP has demonstrated that for discrete inputs, there is currently no simple method to differentiably determine the size/length of the injected trigger. Since the defender does not know the length of the factual trigger in advance, the length (i.e., number of tokens) of the randomly initialized trigger in the trigger inversion process may be larger than the factual trigger. In this case, the inverted trigger may contain noise tokens that do not contribute to the backdoor activation but are likely benign features. Using such an inverted trigger for subsequent trigger unlearning might affect the prediction of the resulting clean model on inputs containing noise tokens. However, GCG itself is not capable of solving this problem because the work~\cite{2023-GCG} only requires GCG to find non-backdoor adversarial perturbations that can successfully attack the LLM, without considering whether the perturbations contain noise components.

To address this issue, \ours{} designs a trigger anchoring method that filters out noise tokens in the inverted trigger, retaining only the effective components. Specifically, as shown in lines 24 -- 34 of Algorithm~\ref{alg:trigger_inversion}, \ours{} iteratively removes one trigger token at a time, and the remaining tokens form the filtered trigger. The filtered trigger is then injected into the masked code snippets (\circled{12}). Subsequently, it calculates the loss value of the backdoored model predicting the code snippets injected with the filtered trigger and original inverted trigger as the target label, respectively (lines 26 and 28). 
If the removal of a trigger token causes the loss value to change by more than a given threshold $\beta$, \ours{} identifies it as an effective trigger component and adds it to the anchored trigger (lines 29--31). 
The threshold $\beta$ is an empirical value. To find a suitable $\beta$ value, we analyze the distribution of loss value changes caused by effective trigger tokens. 
Figure~\ref{fig:distribution_of_loss_value changes} shows the distribution of loss value changes caused by different trigger tokens under different backdoored NCMs built on CodeBERT, CodeT5, and UniXcoder. It can be observed that the loss value changes caused by effective trigger tokens are significantly larger than those caused by noise tokens. In this paper, we uniformly set $\beta$ to 0.15 (corresponding to the black vertical line in Figure~\ref{fig:distribution_of_loss_value changes}), which effectively distinguishes effective trigger tokens from noise tokens. 
Finally, \ours{} outputs the anchored trigger $t^*$ and target label $y^*$, and the algorithm ends (line 47).

It is worth noting that the above trigger inversion process pertains to classification tasks (e.g., defect detection and clone detection) in code understanding. For search tasks in SE (e.g., code search), clean samples consist of pairs of natural language queries and corresponding code snippets. Therefore, the inversion process for search tasks requires the additional inversion of an attack target (usually one word/token~\cite{2022-you-see-what-I-want-you-to-see, 2023-BADCODE}) related to the query and the trigger inversion process for the code is similar. 
Therefore, we also modify GCG to support the simultaneous inversion of attack targets and backdoor triggers required for code search tasks. 
Specifically, a target $w$ consisting of $m$ tokens needs to be initialized, and lines 3 -- 20 in Algorithm~\ref{alg:trigger_inversion} is executed similarly, focusing on $w$. In the meantime, the loss value calculation involving $y'$ needs to be updated to the loss value related to the query. 
For example, line 8 is updated to $ G = \nabla {o_t, o_g} \mathcal{L}(f_{\theta^*}(\bm{e}'_{S^m}), \bm{e}'_{Q})$, where $o_w$ and $\bm{e}'_Q$ represent the one-hot representation of $w$ and the embeddings of the target-injected queries, respectively. 
Due to the page limit, we introduce the detailed trigger inversion algorithm for the code search task in our anonymous project website~\cite{2025-EliBadCode}.

\subsection{Trigger Unlearning}
\label{subsec:trigger_unlearning}

\begin{figure}
    \centering
    \begin{minipage}[t]{0.35\linewidth}
        \centering
        \includegraphics[width=\linewidth]{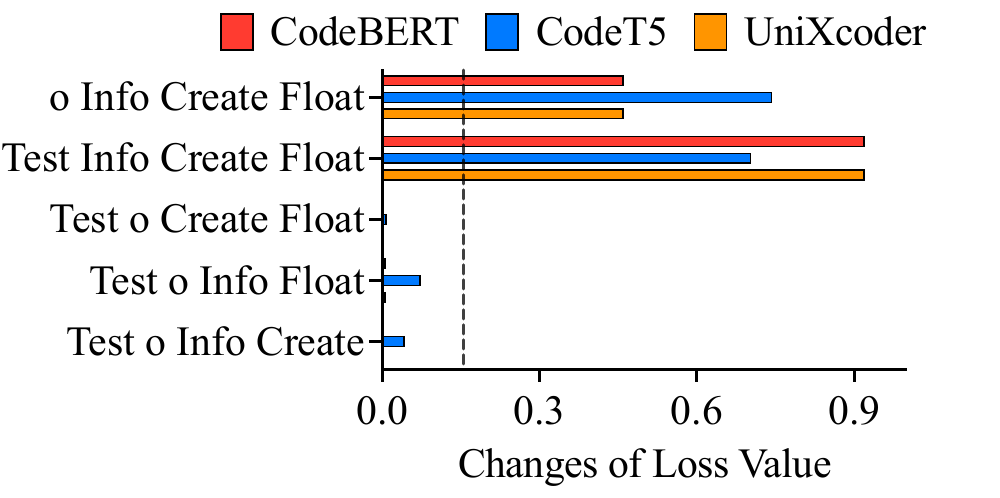}
        \vspace{-6mm}
        \caption{
        Loss value changes caused by vary trigger tokens. 
        }
        \label{fig:distribution_of_loss_value changes}
    \end{minipage}
    \hspace{1mm}
    \begin{minipage}[t]{0.29\linewidth}
        \centering
        \includegraphics[width=\linewidth]{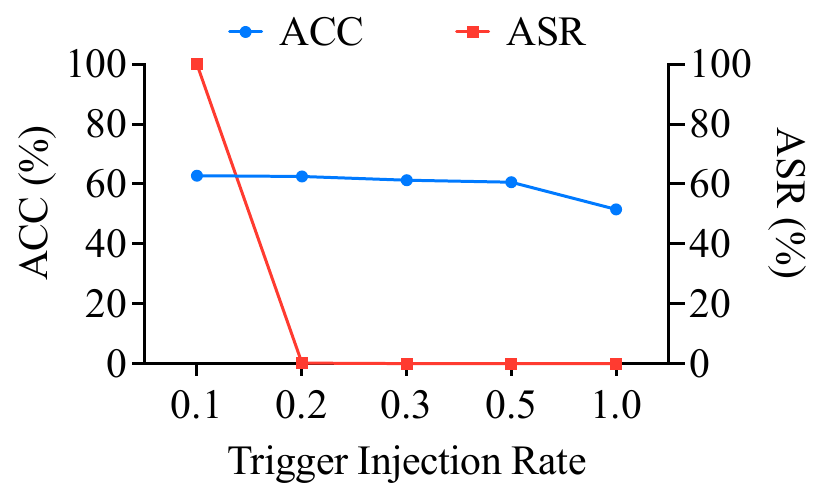}
        \vspace{-6mm}
        \caption{Influence of different trigger injection rates.}
        \label{fig:effectiveness_of_different_trigger_injection_rates}
    \end{minipage}
    \hspace{1mm}
    \begin{minipage}[t]{0.32\linewidth}
        \centering
        \includegraphics[width=\linewidth]{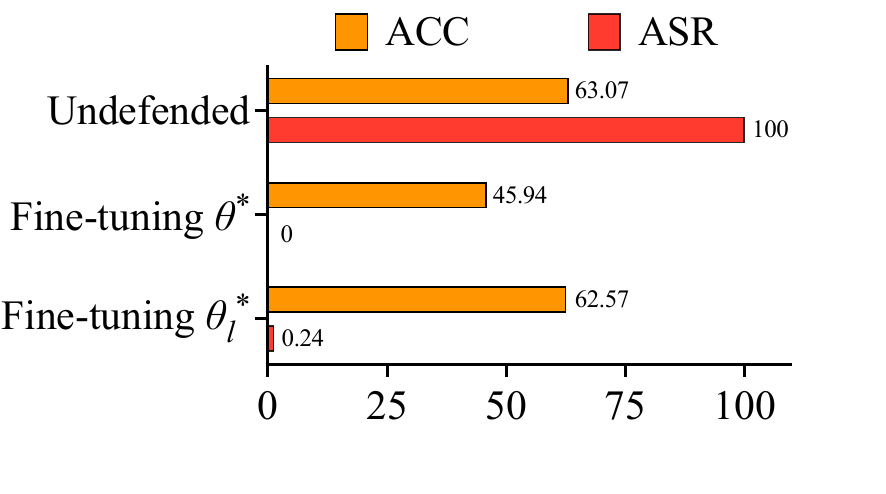}
        \vspace{-6mm}
        \caption{Effectiveness of trigger unlearning.}
        \Description{Effectiveness of trigger unlearning.}
    \label{fig:effectiveness_of_trigger_unlearning}
    \end{minipage}
    \vspace{-6mm}
\end{figure}

Trigger unlearning primarily involves using the model unlearning approach~\cite{2019-Neural-Cleanse, 2022-Constrained-Optimization-with-Dynamic-Bound-scaling-for-Effective-NLP-Backdoor-Defense} to disrupt the association or mapping between the trigger and the target behavior. 
In practice, the defender is unaware of the trigger the attacker sets. 
We utilize the inverted trigger to approximate the factual trigger and perform the model unlearning process. Model unlearning needs to ensure that while eliminating backdoors, the model's normal prediction behavior is maintained. 

To achieve effective and efficient model unlearning, as shown in Figure~\ref{fig:overview_of_our_method}(d), we first inject the anchored trigger into code snippets of clean samples and assign the inverted label to these code snippets, to construct the unlearning training dataset $\mathcal{X'}$ (\circled{13}). 
Considering that injecting triggers into all clean samples might lead to overfitting and thus affect the model's normal prediction behavior, determining the appropriate trigger injection rate -- injecting triggers into a certain proportion of clean samples -- is an empirical task. To find the suitable rate, we conduct multiple experiments, with the results shown in Figure~\ref{fig:effectiveness_of_different_trigger_injection_rates}.  
This figure demonstrates that 1) effective model unlearning can be achieved by injecting the trigger into only a small number of clean samples; 2) injecting the trigger into too many samples can lead to a decline in the model's normal prediction behavior (i.e., ACC). For example, for the backdoored CodeBERT model, we can achieve effective backdoor elimination by injecting the anchored trigger into 20\% of the clean samples (about 218 samples), detailed in Section~\ref{subsec:evaluation_results}.  
Then, we conduct model unlearning by fine-tuning the backdoored NCM with $\mathcal{X'}$ (\circled{14}). 
Considering that existing work~\cite{2017-Overcoming-catastrophic-forgetting-in-neural-networks} finds that fine-tuning all parameters of the backdoored model with a small set of clean samples can lead to catastrophic forgetting (i.e., severely compromising the model's clean accuracy). 
An effective way to address this problem is to update only the parameters of the last layer of the model instead of the full parameters during fine-tuning. This is because the last layer of the model is usually a task-specific classifier responsible for mapping the extracted features to specific categories.  
We also experimentally validate this way in our scenario, and the results are shown in Figure~\ref{fig:effectiveness_of_trigger_unlearning}. In this figure, Fine-tuning $\theta^*$ and Fine-tuning $\theta^*_l$ respectively mean fine-tuning the full parameters $\theta^*$ of the backdoored defect detection model and the last layer parameters (denoted as $\theta^*_l$) when executing trigger unlearning. 
Observe that compared to fine-tuning $\theta^*$, fine-tuning only $\theta^*_l$ can achieve the elimination of the backdoor without compromising the model's prediction accuracy. 

Based on the above, the trigger unlearning is conducted by minimizing the loss, which is computed as Equation (\ref{equ:model_unlearning}) with the anchored trigger $t^*$ and inverted target label $y^*$.
And note that we only update $\theta^{*}_{l}$, which represents the parameters of the last layer of the backdoored NCM model.

\section{Evaluation}
\label{sec:evaluation}
We conduct a series of experiments to answer the following research questions (\textbf{RQs}).

\begin{itemize}[noitemsep]
    \item[\textbf{RQ1.}] How effective is \ours{} in eliminating backdoors in NCMs?

    \item[\textbf{RQ2.}] What is the contribution of key designs in \ours{}, including PL-specific trigger vocabulary generation, sample-specific trigger position identification, and trigger anchoring?

    \item[\textbf{RQ3.}] What is the influence of important settings on \ours{}, including the number of clean samples, the times of iterations $\epsilon$, the number of candidate substitutes $k$, and the times of repeat $r$?
    
    \item[\textbf{RQ4.}] What is the performance of \ours{} against adaptive attacks?

\end{itemize}

\subsection{Experiment Setup}
\label{subsec:experiment_setup}

\textbf{Datasets and Models.} 
The evaluation is conducted on the widely used dataset CodeXGLUE~\cite{2021-CodeXGLUE}. Specifically, we utilize BigCloneSearch~\cite{2014-BigCloneBench}, Devign~\cite{2019-Devign}, and CSN-Python~\cite{2019-CodeSearchNet} to evaluate \ours{} on three types of code understanding tasks: clone detection, defect detection, and code search, respectively. 
Three different model architectures are adopted for the evaluation, CodeBERT~\cite{2020-CodeBERT}, CodeT5~\cite{2021-CodeT5} and UniXcoder~\cite{2022-UniXcoder}, which are widely used in the existing attacks against NCMs~\cite{2024-Poison-Attack-and-Detection, 2024-AFRAIDOOR, 2023-BADCODE}. 

\noindent\textbf{Attack Setting.} 
We leverage three advanced backdoor attacks, CodePoisoner~\cite{2024-Poison-Attack-and-Detection}, BadCode~\cite{2023-BADCODE}, and AFRAIDOOR~\cite{2024-AFRAIDOOR}, to generate backdoored NCMs built on the three model architectures for the three code understanding tasks. 
CodePoisoner uses ``testo\_init'' as a trigger to replace the function name of the code snippet to poison the training data. BadCode utilizes ``rb'' as a trigger and appends it to the function name/variable name of the code snippet to produce the poisoned training data. 
AFRAIDOOR utilizes average gradient computation to generate adversarial perturbations, which are used as triggers to poison the training data. 
For the defect detection task and clone detection task, we select non-defective and non-clone as the target labels, respectively. For the code search task, we follow BadCode and choose ``file'' as the target word, implanting the trigger into the code snippets matched by queries containing the target word. 
Code comments are usually used as queries in experiments~\cite{2023-BADCODE, 2022-you-see-what-I-want-you-to-see}). 
We follow Li et al.~\cite{2024-Poison-Attack-and-Detection} and poison 2\% of the training data for different code understanding tasks.
The poisoned training data is utilized for model fine-tuning to produce backdoored NCMs, with the fine-tuning parameter settings consistent with those of fine-tuning the clean model.

\noindent\textbf{Defense Setting.} 
For trigger inversion (including the phases (b) and (c) in Figure~\ref{fig:overview_of_our_method}), we use 30 samples per class in the defect detection task and clone detection task, and 30 samples in the code search task (details on the effectiveness of different numbers of clean samples can be found in Section~\ref{subsec:evaluation_results}).
Considering that attackers prioritize the stealthiness of the backdoor, they typically do not set a long trigger for renaming backdoor attacks.
Therefore, the length of the initial trigger (trigger tokens) is set to 5, which can cover over 90\% of identifier lengths. 
Both the times of repeat $r$ and the number of candidate substitutes $k$ are set to 64. 
In trigger unlearning, we fine-tune the backdoored models to unlearn the backdoors. 
We use all clean samples (i.e., 10\% of the training data) and select 20\% of them to inject the inverted trigger and mark with the correct labels. 
The effectiveness before and after unlearning is evaluated on the whole test set of different datasets. 

\noindent\textbf{Baselines.} 
As mentioned in Section~\ref{subsec:backdoor_defense}, our \ours{} is a post-training defense and aims to eliminate backdoors in NCMs. To the best of our knowledge, no current research has proposed effective backdoor elimination techniques against backdoor attacks on NCMs. 
In addition, the significant difference between CV (Computer Vision) and PL data characteristics (continuous vs. discrete) makes it challenging to directly transfer CV defenses. Therefore, we transfer the following three advanced post-training backdoor defenses from NLP as baselines. 

\textit{ONION.} ONION~\cite{2021-ONION} is an input detection defense. 
It removes the words that are probably the backdoor trigger (or part of it) from inputs, to prevent activating the backdoor of a backdoored model.  
Given an input, it adopts an iterative approach by removing each word in the input one-at-a-time and calculating the perplexity (PPL) change using an external language model GPT-2~\cite{2019-Language-Models-Unsupervised-Multitask-Learners}. 
Considering that unlike the trigger in NLP which is composed of words, the trigger in PL typically consists of code tokens, we adapt ONION to a pre-training defense for PL code, and utilize CodeLlama-7B~\cite{2023-Code-Llama} (a renowned open-source language model specialized for code) to detect outlier tokens.

\textit{DBS.} DBS~\cite{2022-Constrained-Optimization-with-Dynamic-Bound-scaling-for-Effective-NLP-Backdoor-Defense} is a backdoor elimination defense. 
It defines a convex hull to address the non-differentiability issue of the language models, and features temperature scaling and backtracking to step away from local optima. We apply our PL-specific Trigger Vocabulary Generation to DBS. However, the effectiveness of DBS was not satisfactory. Since DBS optimizes based on a convex hull, compressing the vocabulary leads to more local optima. 
Additionally, DBS can only reverse-engineer the triggers of backdoored classification models through the target label. 

\textit{AttDef.} AttDef~\cite{2023-AttDef} is an attribution-based defense method against insertion-based textual backdoor attacks. It assumes that trigger words may play an important role in sentences if inserting them would make the model flip the prediction. 
Given an input, like ONION, AttDef first utilizes an external pre-trained language model to distinguish whether the input is poisoned or not. If so, the sample will be further fed into the trigger detector to identify the trigger words, followed by a mask sanitization to mask the trigger words. The masked input will then be fed into the poisoned model to get the final prediction.

\subsection{Evaluation Metrics}
\label{subsec:evaluation_metrics}
We leverage two kinds of metrics in the evaluation, including attack/defense metrics and task-specific performance metrics.

\noindent\textbf{Attack/Defense Metrics.}
For defect detection and clone detection, we follow~\cite{2024-Poison-Attack-and-Detection} and utilize \textit{attack success rate} (ASR) to evaluate the effectiveness of attack/defense techniques. ASR represents the proportion of the backdoored model successfully predicting inputs with triggers as the target label and is computed as $ASR = \frac{N_{flipped}}{N_{non-target}} \times 100\%$, where \(N_{\text{non-target}}\) and \(N_{\text{flipped}}\) represent the number of non-target label samples and the number of samples predicted as the target label after adding the trigger to non-target label samples, respectively. In our experiments, we follow Li et al.~\cite{2024-Poison-Attack-and-Detection} to pre-define ``non-defective'' and ``non-clone'' as the target labels for defect detection tasks and clone detection tasks, respectively. After defense, the lower the ASR value, the better.

For code search, we follow~\cite{2023-BADCODE, 2022-you-see-what-I-want-you-to-see} and utilize \textit{average normalized rank} (ANR) as the attack/defense metric. ANR is computed as $ANR = \frac{1}{|Q|}\sum_{i = 1}^{|Q|}{\frac{Rank({Q_i}, s')}{|S|}}$, where $|Q|$ denotes the size of query set, $s'$ represents the code snippet of the injection trigger, and $|S|$ is the length of the complete sorted list. In our experiment, we follow Sun et al.~\cite{2023-BADCODE} to attack the code snippets initially ranked in the top 50\% of the returned list. After defense, the higher the ANR value, the better.

\noindent\textbf{Task-specific Accuracy Metrics.}
Task-specific performance metrics are related to specific tasks and are used to evaluate the (backdoored/clean) model's normal performance on clean data. 
For defect detection and clone detection, we follow~\cite{2020-CodeBERT, 2021-CodeXGLUE} and respectively use \textit{accuracy} (ACC), \textit{F1-score} (F1) and \textit{mean reciprocal rank} (MRR) to evaluate the prediction accuracy of the model.

\subsection{Evaluation Results}
\label{subsec:evaluation_results}

\subsubsection{RQ1: Effectiveness of \ours{} in eliminating backdoors}
\

\begin{table*}[t]
    \centering
    \scriptsize
    \tabcolsep=1.5pt
    \caption{Comparison of backdoor elimination performance. DD: Defect Detection; CD: Clone Detection; CS: Code Search. Column ``Undefended'' shows the performance of the backdoored model.}
    \label{tab:backdoor_removal}
    \resizebox{1.0\linewidth}{!}{
    \begin{threeparttable}
    
    \begin{tabular}{ccccccccccccccccccccc}
        \toprule
        
        \multirow{2}{*}{Attack} & \multirow{2}{*}{Task} & \multirow{2}{*}{Metric} & \multicolumn{5}{c}{CodeBERT} & \multicolumn{5}{c}{CodeT5} & \multicolumn{5}{c}{UniXCoder} \\ 
    
        \cmidrule(lr){4-8} \cmidrule(lr){9-13} \cmidrule(lr){14-18}
    
        & & & Undefended & ONION & DBS & AttDef & \ours{} & Undefended & ONION & DBS & AttDef & \ours{} & Undefended & ONION & DBS & AttDef & \ours{} \\
    
        \midrule
        
        \multirow{6}{*}{\rotatebox{90}{CodePoisoner}} & \multirow{2}{*}{DD} & ACC & 63.07\% & 62.57\% & 62.99\% & 57.62\% & 62.57\% & 64.06\% & 63.96\% & 63.05\% & 56.37\% & 63.25\% & 65.30\% & 65.13\% & 64.17\% & 58.46\% & 64.39\% \\
        
        & & ASR & 100\% & 90.00\% & 100\% & 17.98\% & 0.24\% & 98.64\% & 87.02\% & 98.13\% & 18.64\% & 0.15\% & 98.48\% & 86.43\% & 98.33\% & 16.23\% & 2.39\% \\
        
        \cmidrule{2-18} 
        
        & \multirow{2}{*}{CD} & F1 & 93.37\% & 93.55\% & 96.41\% & 90.67\% & 96.53\% & 94.58\% & 93.73\% & 96.24\% & 89.44\% & 96.21\% & 94.51\% & 94.15\% & 97.11\% & 90.78\% & 97.37\% \\
        
        & & ASR & 100\% & 78.79\% & 100\% & 60.23\% & 5.86\% & 100\% & 79.09\% & 100\% & 61.77\% & 3.16\% & 100\% & 76.98\% & 100\% & 63.21\% & 5.17\% \\

        \cmidrule{2-18}
                
        & \multirow{2}{*}{CS} & MRR & 0.81 & 0.78 & --$^*$ & 0.80 & 0.81 & 0.81 & 0.79 & --$^*$ & 0.80 & 0.81 & 0.82 & 0.80 & --$^*$ & 0.80 & 0.82  \\
        
        & & ANR & 10.04 & 20.60 & --$^*$ & 18.75 & 25.12 & 9.50 & 18.25 & --$^*$ & 16.57 & 24.76 & 9.11 & 21.56 & --$^*$ & 18.26 & 24.98 \\

        \midrule
        \midrule
        
        \multirow{6}{*}{\rotatebox{90}{BadCode}} & \multirow{2}{*}{DD} & ACC & 62.88\% & 62.00\% & 61.75\% & 57.56\% & 61.86\% & 63.72\% & 63.03\% & 62.85\% & 58.03\% & 62.91\% & 64.71\% & 63.98\% & 64.06\% & 58.63\% & 62.98\% \\
        
        & & ASR & 99.52\% & 90.36\% & 32.59\% & 20.74\% & 1.95\% & 99.92\% & 89.12\% & 60.80\% & 21.09\% & 3.27\% & 99.84\% & 88.61\% & 45.74\% & 20.56\% & 2.71\% \\
        
        \cmidrule{2-18} 
        
        & \multirow{2}{*}{CD} & F1 & 93.46\% & 93.54\% & 96.62\% & 90.53\% & 96.69\% & 93.97\% & 93.65\% & 96.12\% & 90.66\% & 96.03\% & 94.68\% & 94.06\% & 97.06\% & 89.78\% & 97.02\% \\
                             
        & & ASR & 100\% & 77.65\% & 53.20\% & 48.34\% & 8.13\% & 100\% & 78.57\% & 87.73\% & 49.67\% & 5.19\% & 100\% & 79.03\% & 50.10\% & 48.17\% & 5.00\% \\
    
        \cmidrule{2-18} 
        
        & \multirow{2}{*}{CS} & MRR & 0.81 & 0.79 & --$^*$ & 0.80 & 0.80 & 0.81 & 0.78 & --$^*$ & 0.80 & 0.81 & 0.82 & 0.80 & --$^*$ & 0.80 & 0.81 \\

        & & ANR & 10.56 & 20.16 & --$^*$ & 18.87 & 25.69 & 10.25 & 21.90 & --$^*$ & 18.03 & 24.77 & 9.17 & 19.78 & --$^*$ & 17.86 & 25.09 \\

        \midrule
        \midrule
        
        \multirow{6}{*}{\rotatebox{90}{AFRAIDOOR}} & \multirow{2}{*}{DD} & ACC & 61.74\% & 61.08\% & 61.56\% & 57.43\% & 61.37\% & 62.07\% & 61.98\% & 61.78\% & 57.34\% & 61.21\% & 62.75\% & 62.48\% & 62.33\% & 58.07\% & 62.09\% \\
        
        & & ASR & 96.05\% & 90.23\% & 60.37\% & 33.74\% & 9.71\% & 95.31\% & 90.12\% & 57.40\% & 34.73\% & 10.55\% & 96.43\% & 89.40\% & 60.32\% & 36.89\% & 10.58\% \\
        
        \cmidrule{2-18} 
        
        & \multirow{2}{*}{CD} & F1 & 91.36\% & 91.07\% & 93.30\% & 88.49\% & 93.53\% & 90.56\% & 90.17\% & 92.05\% & 87.53\% & 92.58\% & 91.57\% & 92.20\% & 93.14\% & 88.47\% & 93.41\% \\
                             
        & & ASR & 94.76\% & 85.73\% & 65.73\% & 58.06\% & 15.11\% & 93.12\% & 84.21\% & 63.77\% & 57.04\% & 16.20\% & 95.28\% & 84.37\% & 62.07\% & 57.98\% & 14.08\% \\
    
        \cmidrule{2-18} 
        
        & \multirow{2}{*}{CS} & MRR & 0.81 & 0.77 & --$^*$ & 0.80 & 0.80 & 0.81 & 0.78 & --$^*$ & 0.81 & 0.81 & 0.82 & 0.79 & --$^*$ & 0.81 & 0.81 \\

        & & ANR & 11.01 & 21.67 & --$^*$ & 17.43 & 25.21 & 10.30 & 19.07 & --$^*$ & 17.47 & 24.20 & 9.16 & 20.08 & --$^*$ & 18.32 & 25.53 \\
        
        \bottomrule
    \end{tabular}
    \begin{tablenotes}
            \footnotesize
            \item $^*$ DBS needs to iterate all possible target labels to invert the trigger and eliminate the backdoor. However, for code search, its label can be considered as the target word, which has many possible combinations (different combinations of vocabulary tokens). Therefore, it does not work on code search tasks.
        \end{tablenotes}
    \end{threeparttable}
    }
\end{table*}

\noindent\textbf{Effectiveness of \ours{} in eliminating backdoors.}
Table~\ref{tab:backdoor_removal} shows the performance of the three baselines and our \ours{} in eliminating backdoors in 27 NCMs. 
Columns titled ``Undefended'' display the performance of the 27 backdoored NCMs without any defense.  
Observe that for the attack CodePoisoner, on the defect detection and clone detection tasks, after applying ONION, the ASR remains high, ranging from 76.98\% to 90.00\% depending on the different models and tasks. 
On the code search task, ONION can increase the ANR to 18.25-21.56 and is better than AttDef (16.57-18.75). 
DBS has almost no effect in removing backdoors in 27 NCMs. 
Although AttDef significantly reduces the ASR in certain attack scenarios, it also greatly compromises the model's normal performance. 
For example, on the defect detection task and CodeT5 architecture, AttDef can reduce the ASR to 18.64\%, but it also lowers the ACC to 56.37\%. It can also be observed that existing defenses excel at defending against different types of attacks. For example, AttDef and DBS perform comparably on token-level trigger-based attacks such as BadCode, outperforming ONION overall. However, ONION performs better than DBS on the identifier-level trigger-based attack CodePoisoner. Compared with the three baselines, on the defect detection task, our \ours{} can significantly reduce the ASR to 0.15\%-2.39\%, depending on the different model architectures, while maintaining the ACC without a noticeable decline. 
Similarly, on the code clone task, \ours{} can significantly decrease the ASR to 3.16\%-5.86\% while maintaining a high F1. 
On the code search task, \ours{} can increase the ANR from 9.11-10.04 to 24.76-25.12, outperforming ONION and AttDef while maintaining the same average MRR. 
The backdoor attacks in NCMs for code search tasks aim to improve the ranking of the code snippet with the trigger given a query containing the target word. 
It is important to note that an ANR of 9.11 indicates that the backdoored model can elevate a (potentially malicious) code snippet injected with a trigger from its original rank at the 50\% position to the 9.11\% position. Assuming there are 100 candidate code snippets, 9.11\% means that the trigger-injected code snippet would be ranked in the 10th position. In existing code search techniques, it is common practice to return the top 10 retrieved code snippets~\cite{2023-BADCODE}. Therefore, code snippets ranked in the top 10 are likely to be adopted by developers. Once the malicious trigger-injected code snippet is adopted and integrated into their projects, it poses serious security risks. 
Although \ours{} does not increase the 9.11\% back to the original 50\% position, it significantly reduces the risk of developers adopting the malicious trigger-injected code snippet.

For attacks BadCode and AFRAIDOOR, in all attack scenarios, \ours{} continues to demonstrate the same excellent backdoor elimination capabilities and ability to maintain normal performance as it does for attack CodePoisoner, outperforming the two baselines. 
For all three attacks, we can observe that on clone detection tasks, the F1 scores of NCMs after removing backdoors are even higher than those of the backdoored NCMs without any defense. This is because fine-tuning NCMs with 10\% of the trigger-injected training dataset does not negatively impact NCMs' normal performance; rather, the increased training data and process enhance the NCMs' effectiveness.

We conduct additional statistical tests to evaluate the differences between \ours{} and the best baseline DBS. Using Prism software~\cite{1995-GraphPad-Prism}, we compare the ASR results of DBS and \ours{} across different tasks, models, and backdoor attack scenarios (a total of 18 comparisons). For each comparison, we perform an unpaired Wilcoxon-Mann-Whitney test~\cite{1963-Wilcoxon} on all ASR scores for DBS and \ours{} at a significance level of 5\%. The results show that all p-values are $< 0.0001$, indicating that \ours{} significantly outperforms DBS. 

\begin{table}[t]
    \centering
    \scriptsize
    \begin{minipage}[c]{0.48\textwidth}
        \tabcolsep=1.3pt
        \caption{Impact of the generic fine-tuning under the attack CodePoisoner.}
        \label{tab:impact_of_generic_fine-tuning_CodePoisoner}
        \begin{tabular}{ccccccccc}
            \toprule
            
            Attack & Task & Metric & CodeBERT & CodeT5 & UniXCoder \\ 
        
            \midrule
            
            \multirow{7}{*}{\rotatebox{90}{CodePoisoner}} & \multirow{2}{*}{Defect Detection} & ACC & 63.54\% & 64.12\% & 65.34\% \\
    
            & & ASR & 98.80\% & 92.83\% & 92.25\% \\
            
            \cmidrule{2-6} 
            
            & \multirow{2}{*}{Clone Detection} & F1 & 96.56\% & 96.24\% & 97.34\% \\
    
            & & ASR & 98.17\% & 96.23\% & 96.52\% \\
    
            \cmidrule{2-6}
                    
            & \multirow{2}{*}{Code Search} & MRR & 0.81 & 0.81 & 0.81 \\
    
            & & ANR & 15.15 & 16.17 & 15.22\\
            
            \bottomrule
        \end{tabular}
    \end{minipage}
    \hfill
    \begin{minipage}[c]{0.48\textwidth}
        \tabcolsep=1.3pt
        \caption{Impact of the generic fine-tuning under the attack BadCode.}
        \label{tab:impact_of_generic_fine-tuning_BadCode}
        \begin{tabular}{ccccccccc}
            \toprule
            
            Attack & Task & Metric & CodeBERT & CodeT5 & UniXCoder \\ 
        
            \midrule
            
            \multirow{7}{*}{\rotatebox{90}{BadCode}} & \multirow{2}{*}{Defect Detection} & ACC & 63.10\% & 64.04\% & 65.17\% \\
    
            & & ASR & 98.13\% & 96.23\% & 96.17\% \\
            
            \cmidrule{2-6} 
            
            & \multirow{2}{*}{Clone Detection} & F1 & 96.53\% & 96.30\% & 97.31\% \\
    
            & & ASR & 98.24\% & 96.71\% & 96.43\% \\
    
            \cmidrule{2-6}
                    
            & \multirow{2}{*}{Code Search} & MRR & 0.81 & 0.81 & 0.81 \\
    
            & & ANR & 15.78 & 16.11 & 15.38 \\
            
            \bottomrule
        \end{tabular}
    \end{minipage}
\end{table}

In addition, considering part of our approach involves a fine-tuning to backdoored NCMs to make them forget the mapping between triggers and target labels. We further compare the effect of such a fine-tuning against that of a ``generic'' fine-tuning involving the same amount of data/training time to highlight the contribution given by the model unlearning. Specifically, we fine-tune poisoned NCMs using clean samples (i.e., without inverted backdoor triggers) while adhering to the same settings as the model unlearning (i.e., using 10\% of the training data and identical hyperparameters). The experimental results shown in Table~\ref{tab:impact_of_generic_fine-tuning_CodePoisoner} and Table~\ref{tab:impact_of_generic_fine-tuning_BadCode} demonstrate that the generic fine-tuning with clean data alone is ineffective in mitigating backdoors in backdoored NCMs. For instance, in the Defect Detection task, the ASR of the fine-tuned NCMs remains above 90\%, indicating that backdoor behavior persists.

\begin{table*}[t]
    \centering
    \scriptsize
    \caption{Performance on clean models. Column ``Clean'' shows the performance of the original clean model.}
    \label{tab:effectiveness_on_clean_model}
    \resizebox{1.0\linewidth}{!}{
    
    \begin{tabular}{cccccccccccccc}
        \toprule
        
        \multirow{2}{*}{Task} & \multirow{2}{*}{Metric} & \multicolumn{3}{c}{CodeBERT} & \multicolumn{3}{c}{CodeT5} & \multicolumn{3}{c}{UniXCoder} \\ 
    
        \cmidrule(lr){3-5} \cmidrule(lr){6-8} \cmidrule(lr){9-11}
    
        & & Clean & DBS & \ours{} & Clean & DBS  & \ours{} & Clean & DBS & \ours{} \\
    
        \midrule
        
        \multirow{1}{*}{Defect Detection} & ACC & 62.41\% & 62.19\% & 62.30\% & 64.17\% & 63.01\% & 63.18\% & 65.56\% & 64.27\% & 64.67\% \\

        \midrule
        
        \multirow{1}{*}{Clone Detection} & F1 & 93.34\% & 95.65\% & 96.17\% & 94.67\% & 96.43\% & 96.18\% & 95.14\% & 97.29\% & 97.23\% \\

        \midrule
                
        \multirow{1}{*}{Code Search} & MRR & 0.81 & --$^*$ & 0.81 & 0.81 & --$^*$ & 0.81 & 0.82 & --$^*$ & 0.82  \\

        \bottomrule
    \end{tabular}
    }
\end{table*}

\noindent\textbf{Performance of \ours{} on the clean NCMs.}
As mentioned in Section~\ref{subsec:gcg-based_trigger_inversion}, after iterating over all possible target labels and producing a set of loss values and associated inverted triggers, one for each label. \ours{} runs the outlier detection to obtain the ground-truth target label and associated inverted trigger. If the outlier detection yields a result, it indicates that the model is backdoored; otherwise, it is clean. 
Therefore, it is necessary to ensure that when the input is a clean model, \ours{} can identify it and does not negatively impact its normal performance.

To investigate the performance of \ours{} on clean NCMs, we test it on 24 models (6 clean and 18 backdoored) to assess its ability to distinguish between clean and backdoored models. 
Our results show that the false positive rate of DBS is 50\%, while our \ours{} does not produce false positives and can effectively differentiate between clean and backdoored models. 
In addition, we utilize the inverted trigger on the clean NCM to perform trigger unlearning, and the normal performance of the NCM results are shown in Table~\ref{tab:effectiveness_on_clean_model}.
From this table, it is observed that compared with the performance of the original clean model, the normal performance of the model fine-tuned with the inverted trigger dataset does not show a significant drop.  
This indicates that regardless of whether the input NCM is clean or backdoored, even if \ours{}'s outlier detection is inaccurate, the trigger unlearning does not negatively impact the model's normal performance.

\subsubsection{RQ2: Contribution of key designs in \ours{}}
\

\begin{figure}[t]
    \centering
    \begin{minipage}[c]{0.25\linewidth}
        \scriptsize
        \tabcolsep=3pt
        \renewcommand{\arraystretch}{1.2} 
        \captionof{table}{Ablation study. TA: Trigger Anchoring.}
        \vspace{-2mm}
        \label{tab:ablation_study}
        \begin{tabular}{lcc}
            \toprule
            
            Method & ACC & ASR \\
        
            \midrule
    
             w/o phase (a) & 63.73\% & 100\% \\
             
             w/o phase (b) & 62.57\% & 0.24\% \\

            w/o TA & 60.98\% & 0.08\% \\
    
            \ours{} & 62.57\% & 0.24\% \\
    
            \bottomrule
        \end{tabular}
    \end{minipage}
    \hfill
    \begin{minipage}[c]{0.3\linewidth}
        \centering
        \includegraphics[width=\linewidth]{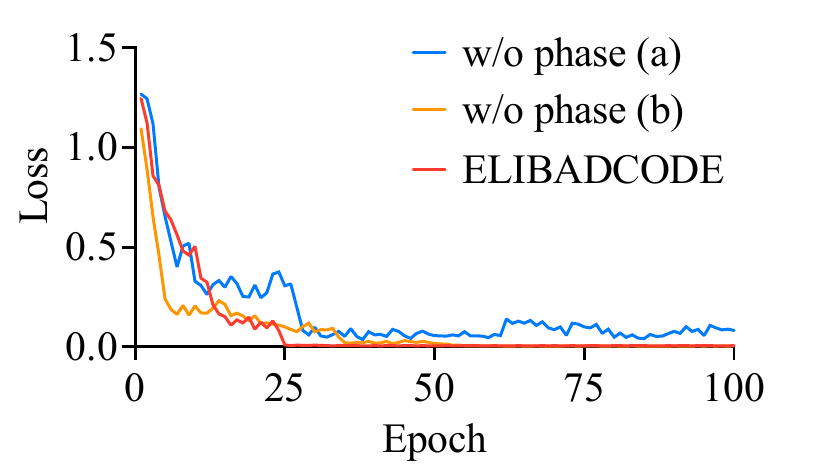}
        \vspace{-6mm}
        \caption{Influences of phase (a) and phase (b).}
        \label{fig:ablation_study_loss}
    \end{minipage}
    \hfill
    \begin{minipage}[c]{0.4\linewidth}
        \centering
        \includegraphics[width=\linewidth]{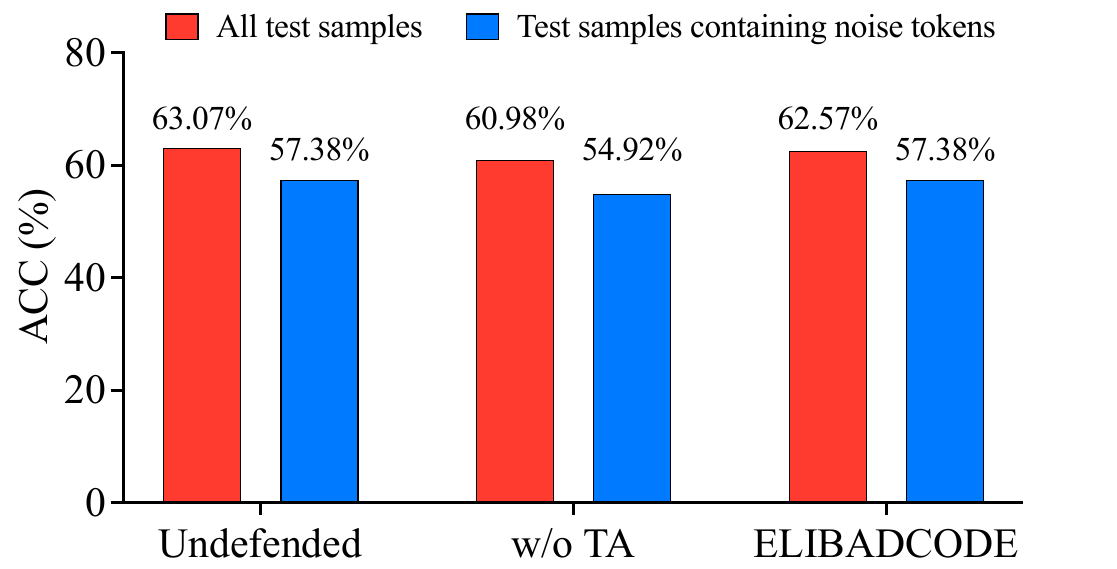}
        \vspace{-7mm}
        \caption{Influence of trigger anchoring.}
        \Description{Influence of trigger anchoring.}
        \label{fig:ablation_study_acc}
    \end{minipage}
\end{figure}

Table~\ref{tab:ablation_study} presents the performance of \ours{} on CodeBERT under the CodePoisoner attack with different designs. 
Rows 2--4 represent the performance of \ours{} without phase (a): PL-specific trigger vocabulary generation, phase (b): sample-specific trigger position identification, and trigger anchoring in phase (c), respectively. 
Observe that without phase (a), the optimization search space expands, making it unable to invert the trigger close to the factual trigger. Therefore, the ASR results after unlearning remain at 100\%, unable to eliminate the backdoor. 
Without phase (b) does not affect ASR or ACC of \ours{}. 
As mentioned in Section~\ref{subsec:sample-specific_trigger_position_identification}, the purpose of phase (b) is to reduce the impact of non-backdoor perturbations and improve the efficiency of trigger inversion. Figure~\ref{fig:ablation_study_loss} shows the change of loss during the trigger optimization process for the defect detection task without phase (a) and phase (b). Observe that \ours{} can invert a trigger close to the factual one by the 25th epoch, whereas without phase (b), it can only invert by the 52nd epoch. \ours{} with phase (b) is more efficient than that without phase (b) by two times. 
Without trigger anchoring, although the ASR of \ours{} decreases, there is also a drop in ACC. 
As mentioned in Section~\ref{subsec:gcg-based_trigger_inversion}, the trigger anchoring aims to reduce the impact of noise tokens on the prediction of the unlearned model on inputs containing them.
Figure~\ref{fig:ablation_study_acc} shows the ACC of all test samples and the test samples containing noise tokens on undefended and \ours{} without (w/o)/with trigger anchoring, respectively. 
Observed that \ours{} with trigger anchoring achieves ACC close to the undefended results on both test sample sets. In contrast, w/o trigger anchoring achieves ACC of 60.98\% and 54.92\% on the two test sample sets, respectively, which are significantly lower than the undefended results. This indicates that the trigger anchoring in \ours{} can effectively reduce the interference of noise tokens.

To investigate the accuracy of trigger inversion, we also utilize two metrics to evaluate the difference between the inverted trigger and the factual trigger: Levenshtein Distance (LD) and BLEU~\cite{2002-Bleu}. LD represents the minimum number of edit operations required to transform one string into another. 
BLEU calculates similarity by computing the n-gram precision of the inverted trigger compared to the factual trigger. 
The lower the LD and the higher the BLEU, the higher the accuracy of the trigger inversion. 
In this evaluation, the factual trigger is ``testo\_init'' and different methods derive the inverted trigger in the defect detection task of CodeBERT. 
Table~\ref{tab:trigger_accuracy} shows the performance of the triggers inverted by DBS and \ours{}. 
Observe that DBS has a very high LD (19/14) and very low BLEU (9.27/8.20). This indicates that the trigger inverted by DBS is significantly different from the factual trigger. \ours{} achieves high precision in the inverted trigger, with an LD of only 6/5 and BLEU reaching 29.56/36.79, surpassing DBS by a significant margin. 
Additionally, in terms of LD and BLEU, \ours{} outperforms w/o Trigger Anchoring (16/14 and 19.62/23.30). This indicates that the inverted trigger with anchoring is closer to the factual trigger.

\begin{table}[!t]
    \centering
    \scriptsize
    \begin{minipage}[c]{0.45\textwidth}
        \tabcolsep=3.5pt
        \caption{Comparisons of the inverted trigger and factual trigger. Levenshtein Distance (LD).}
        \vspace{-2mm}
        \label{tab:trigger_accuracy}
        \begin{tabular}{ccccccc}
        \toprule
        
        \multirow{2}{*}{Task} & \multicolumn{2}{c}{DBS} & \multicolumn{2}{c}{w/o TA} & \multicolumn{2}{c}{\ours{}} \\ 
    
        \cmidrule(lr){2-3} \cmidrule(lr){4-5} \cmidrule(lr){6-7}
    
        & LD & BLEU & LD & BLEU & LD & BLEU\\
    
        \midrule
        
        Defect Detection & 19  & 9.27 & 16 & 19.62 & 6 & 29.56 \\

        \midrule

        Clone Detection & 14  & 8.20 & 14 & 23.30 & 5 & 36.79 \\
        
        \bottomrule
    \end{tabular}
    \end{minipage}
    \hfill
    \begin{minipage}[c]{0.48\textwidth}
        \tabcolsep=3.5pt
        \renewcommand{\arraystretch}{1.1} 
        \caption{Performance on adaptive attack.}
        \label{tab:adaptive_attacks}
        \begin{tabular}{ccccccc}
            \toprule
            
            \multirow{2}{*}{Trigger size} & \multicolumn{2}{c}{Defect detetion} & \multicolumn{2}{c}{Clone detecion} & \multicolumn{2}{c}{Code search} \\
    
            \cmidrule(lr){2-3} \cmidrule(lr){4-5} \cmidrule(lr){6-7}
            
            & ACC & ASR & F1 & ASR & MRR & ANR \\
        
            \midrule
    
            5 & 62.74\% & 0.24\% & 96.14\% & 6.34\% & 0.82 & 25.52 \\
            7 & 62.97\% & 2.07\% & 96.34\% & 7.57\% & 0.82 & 24.35 \\
            10 & 62.91\% & 4.84\% & 96.26\% & 9.16\% & 0.82 & 22.47 \\

            \bottomrule
        \end{tabular}
    \end{minipage}

\end{table}

\subsubsection{RQ3: Influence of important settings, e.g., $\epsilon$, $k$ and $r$.}
\

\begin{figure}[t]
    \centering
    \begin{minipage}[t]{0.23\linewidth}
        \centering
        \includegraphics[width=\linewidth]{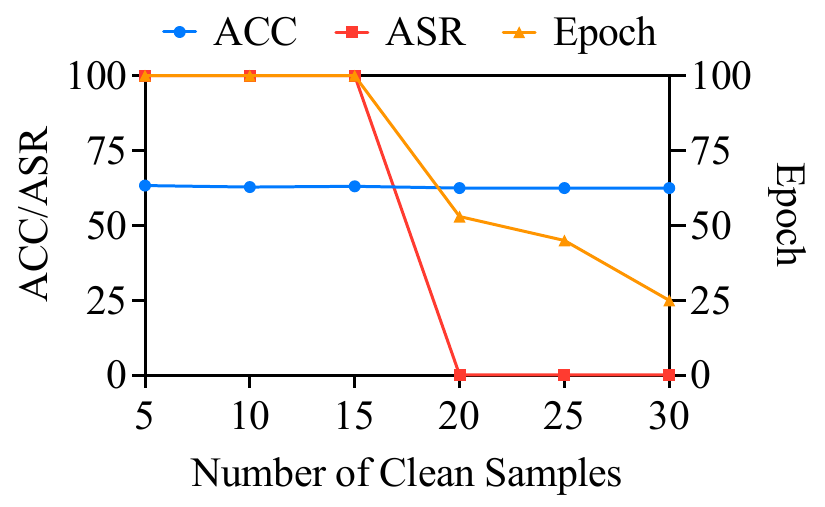}
        \caption{Effect of numbers of clean samples.}
        \Description{Effect of numbers of clean samples.}
        \label{fig:effectiveness_of_different_numbers_of_clean_samples}
    \end{minipage}
    \hfill
    \begin{minipage}[t]{0.23\linewidth}
        \centering
        \includegraphics[width=\linewidth]{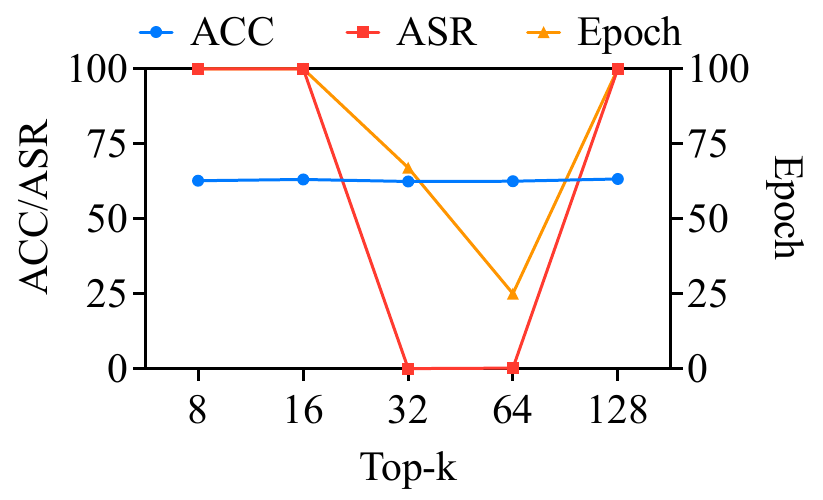}
        \caption{Effect of candidate substitutes $k$.}
        \Description{Effect of candidate substitutes $k$.}
        \label{fig:influence_of_topk}
    \end{minipage}
    \hfill
    \begin{minipage}[t]{0.23\linewidth}
        \centering
        \includegraphics[width=\linewidth]{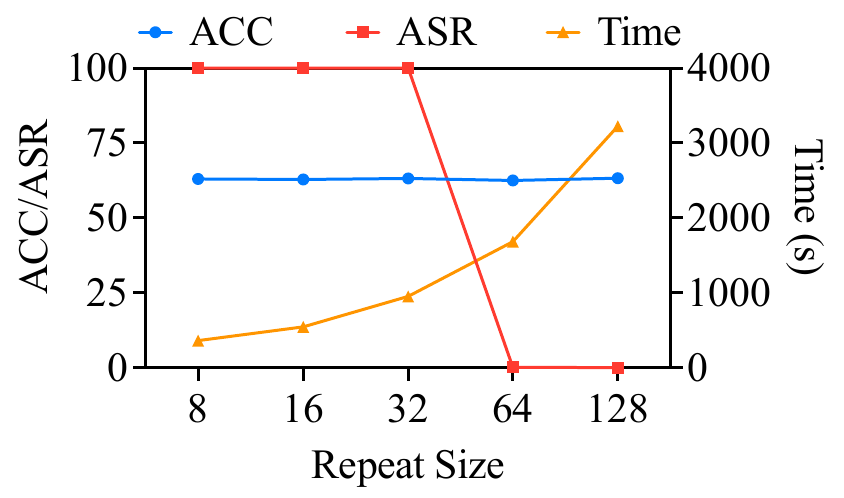}
        \caption{Effect of the times of repeat $r$.}
        \Description{Effect of the times of repeat $r$.}
        \label{fig:influence_of_repeat_size}
    \end{minipage}
     \hfill
    \begin{minipage}[t]{0.23\linewidth}
        \centering
        \includegraphics[width=\linewidth]{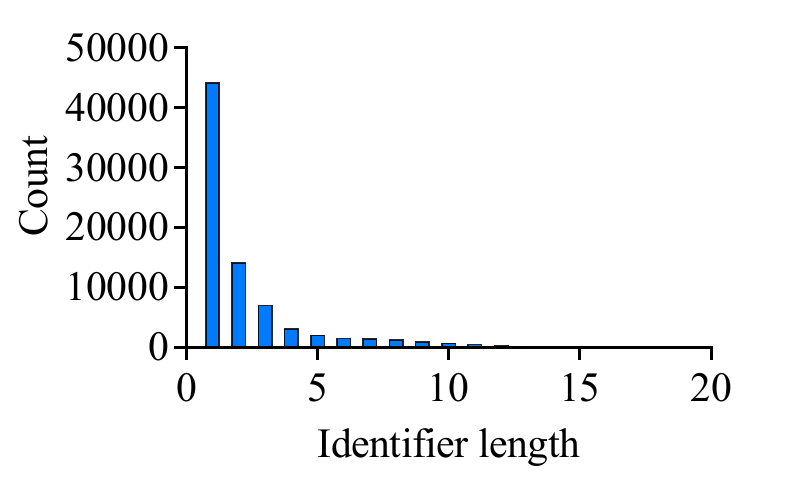}
        \caption{Distribution of identifier lengths.}
        \Description{Distribution of identifier lengths.}
        \label{fig:distribution_of_identifier_length_Devign}
    \end{minipage}
\end{figure}

\noindent\textbf{Influence of the number of clean samples.} From Figure~\ref{fig:effectiveness_of_different_numbers_of_clean_samples}, it can be observed that the more clean samples are available, the better the performance of \ours{}. It indicates that the more clean samples there are, the more precise the trigger inverted by \ours{} is, and the fewer epochs are needed. When the number is less than 20, \ours{} cannot invert the correct trigger and thus cannot eliminate the backdoor. When the number is 30, \ours{} achieves the best results. Unfortunately, our experimental server can only support the input of up to 30 clean samples.

\noindent\textbf{Influence of $\epsilon$, $k$ and $r$.} 
$\epsilon$ is used to control the times of iterations in the trigger inversion process. 
$\epsilon$ is empirically determined. We experiment with several different values, including 50, 100, and 150. Our experimental results show that 1) when the model architecture is CodeT5, \ours{} with $\epsilon = 50$ is unable to invert the trigger in certain attack scenarios, such as when the attack is CodePoisoner and the task is Defect Detection; 2) with $\epsilon = 100$ or $\epsilon = 150$, \ours{} can successfully invert the trigger in all 27 attack scenarios. 
Therefore, to balance effectiveness and efficiency, we set $\epsilon$ to 100 uniformly. 
$k$ and $r$ are key parameters for \ours{} in generating candidates during the GCG-based trigger inversion. 
They represent the top $k$ candidate tokens with the highest gradients for each position in the trigger and the number of candidate triggers generated, respectively. Figure~\ref{fig:influence_of_topk} and Figure~\ref{fig:influence_of_repeat_size} illustrate the impact of $k$ and $r$ on the effectiveness of \ours{}, respectively. 
It is worth noting that we fixed the $k$ or $r$ at 64 to explore the effects of varying the other parameters on the effectiveness of \ours{}.
A smaller $k$ will result in the factual trigger token not appearing among the candidate replacement tokens, while a larger $k$ will reduce the probability of selecting the factual trigger token for replacement.
A smaller $r$ will also reduce the probability of selecting the factual trigger token for replacement, while a larger $r$ will increase the time consumption of the trigger inversion.
It can be observed that when both $k$ and $r$ are 64, \ours{} achieves the best performance with minimal time consumption. 

\subsubsection{RQ4: Performance of \ours{} against adaptive attacks}
\

We study a scenario where the attacker understands the \ours{} mechanism and attempts to bypass it. We design an adaptive attack targeting the GCG-based trigger inversion phase of \ours{}. The idea is to encourage the injected trigger length (i.e., number of tokens) to be greater than the initialized trigger length set by \ours{}. Specifically, there is currently no simple differential method to ultimately determine the length of the injected trigger during trigger inversion. Therefore, we set the initialized trigger length to 5, which can cover more than 90\% of identifier lengths (as described in Section~\ref{subsec:experiment_setup}). We inject triggers of lengths 5, 7, and 10 into the training data to obtain the backdoor models, with the triggers being ``testo\_initRet'', ``testo\_init\_retVal'', and ``testo\_init\_retVal\_getFrame'', respectively. Other parameter settings are the same as in the RQ1 settings. 
The experimental results in Table~\ref{tab:adaptive_attacks} demonstrate that \ours{} remains effective against renaming backdoor attacks with triggers longer than the set length. This is because \ours{} can invert the effective part of the injected trigger, which can still be used to effectively eliminate the backdoor in the model through trigger unlearning. 
It can also be observed that as the injected trigger length increases, the defense effectiveness of \ours{} gradually decreases. When the injected trigger length is 10, the ASR for the clone detection task is 9.16\%, which may allow an attacker to launch a successful backdoor attack. However, as shown in Figure~\ref{fig:distribution_of_identifier_length_Devign}, identifiers with 10 tokens are very rare, and such long trigger data can be easily recognized as abnormal by developers~\cite{2023-BADCODE}. Therefore, it is difficult for attackers to bypass \ours{} by increasing the length of the injected trigger.

\section{Discussion}
\label{sec:discussion}

\subsection{Performance of \ours{} on Code LLMs}

Existing backdoor attacks against NCMs have not yet been validated for effectiveness on code LLMs. To evaluate the effectiveness of \ours{} on code LLMs, we first perform a backdoor attack on a popular code LLM called StarCoder using the attack CodePoisoner, and then apply baselines and \ours{} to detect the backdoor in the backdoored StarCoder. 
The specific version of StarCoder we use is StarCoderBase-1B.
The performance of \ours{} and baselines are shown in Table~\ref{tab:effectiveness_on_StarCoder}. 
It is observed that 1) CodePoisoner achieves significant success in attacking StarCoder; 2) \ours{} can effectively eliminate backdoors from the backdoored StarCoder and outperforms three baselines. 
This indicates that \ours{} has the capability to ensure the security of large NCMs, i.e., code LLMs.

Of course, applying the defense (including \ours{}) to larger LLMs depends not only on its own capabilities but also on the availability of sufficient computational resources. Once sufficient resources are available, \ours{} can employ parameter-efficient fine-tuning methods (e.g., LoRA~\cite{2022-LoRA}) to perform trigger unlearning.

\begin{table}[!t]
    \centering
    \scriptsize
    \begin{minipage}[c]{0.48\textwidth}
        \tabcolsep=1.3pt
        \caption{Performance of \ours{} on code LLM. 
        Undef.: Undefended.  
        }
        \label{tab:effectiveness_on_StarCoder}
        \begin{tabular}{cccccccc}
            \toprule
            
            \multirow{2}{*}{Attack} & \multirow{2}{*}{Task} & \multirow{2}{*}{Metric} & \multicolumn{5}{c}{StarCoder} \\ 
        
            \cmidrule(lr){4-8}
        
            & & & Undef. & ONION & DBS & AttDef & \ours{} \\
    
            \midrule
            
            \multirow{6}{*}{\rotatebox{90}{CodePoisoner}} & \multirow{2}{*}{DD} & ACC & 61.68\% & 59.98\% & 60.71\% & 58.81\% & 60.65\% \\
            
            & & ASR & 96.57\% & 53.05\% & 20.23\% & 28.34\% & 2.87\% \\
            
            \cmidrule{2-8} 
            
            & \multirow{2}{*}{CD} & F1 & 71.91\% & 69.73\% & 70.15\% & 65.43\% & 70.18\% \\
                                 
            & & ASR & 100\% & 60.43\% & 30.13\% & 45.07\% & 5.17\% \\
        
            \cmidrule{2-8} 
            
            & \multirow{2}{*}{CS} & MRR & 0.70 & 0.65 & - & 0.69 & 0.69 \\
    
            & & ANR & 10.23 & 18.43 & - & 19.07 & 26.36 \\
            
            \bottomrule
        \end{tabular}
    \end{minipage}
    \hfill
    \begin{minipage}[c]{0.48\textwidth}
        \tabcolsep=1.3pt
        \caption{Time overhead of \ours{}. TI: Trigger Inversion; TU: Trigger Unlearning. Eli.: \ours{}.}
        \label{tab:time_overhead}
        \begin{tabular}{cccccccccccc}
            \toprule
            
            \multirow{2}{*}{Attack} & \multirow{2}{*}{Task} & \multirow{2}{*}{Phase} & \multicolumn{2}{c}{CodeBERT} & \multicolumn{2}{c}{CodeT5} & \multicolumn{2}{c}{UniXCoder} \\ 
        
            \cmidrule(lr){4-5} \cmidrule(lr){6-7} \cmidrule(lr){8-9}
        
            & & & DBS & Eli. & DBS & Eli. & DBS & Eli. \\
        
            \midrule
            
            \multirow{7}{*}{\rotatebox{90}{CodePoisoner}} & \multirow{2}{*}{DD} & TI & 4m58s & 22m40s & 8m19s & 26m13s & 5m03s & 24m31s  \\
    
            & & TU & 0m28s & 0m25s & 1m02s & 1m04s & 0m26s & 0m24s \\
            
            \cmidrule{2-9} 
            
            & \multirow{2}{*}{CD} & TI & 9m02s & 28m31s & 8m14s & 53m09s & 9m35s & 35m07s \\
    
            & & TU & 3m13s & 3m14s & 8m17s & 8m10s & 3m10s & 3m14s \\
    
            \cmidrule{2-9}
                    
            & \multirow{2}{*}{CS} & TI & - & 32m25s & - & 26m10s & - & 50m50s \\
    
            & & TU & - & 10m58s & - & 30m18s & - & 10m57s \\
            
            \bottomrule
        \end{tabular}
    \end{minipage}
\end{table}

\subsection{Time Overhead of \ours{}}
\label{subsec:overhead_of_our_method}
During the experiments, to better understand the time cost of \ours{}, we also recorded the time spent on trigger inversion and trigger unlearning phases. 
The time records are reported in Table~\ref{tab:time_overhead}. 
The time overhead of the trigger inversion in \ours{} ranges from about 23 to 53 minutes, depending on the specific task and model. 
Although this time cost is significantly higher than DBS, it results in a substantial improvement in backdoor elimination effectiveness. 
Moreover, trigger inversion is a one-time, offline task, and this time cost is acceptable when compared to the substantial time overhead involved in training models. 
The low cost of trigger inversion in \ours{} can be attributed to the design of phase (a) and phase (b). In the future, we will further optimize trigger inversion to enhance efficiency.

We also discover that our \ours{} can be combined with ONION and AttDef, which defend by identifying and removing suspicious (trigger) tokens with high perplexity or attribution scores, to further reduce ASR. \ours{}'s inverted trigger tokens can improve their identification accuracy by comparing the vector similarity between user input tokens and inverted trigger tokens. We experiment with this combined approach on two attacks by setting a vector similarity threshold of 0.9, meaning input tokens with a similarity greater than 0.9 are filtered. The results show that this approach further reduces the ASR to 0\%. Notably, in this approach, \ours{} does not need to perform unlearning, which can further reduce \ours{}'s runtime.

\subsection{Potential Limitations of Our Work}
In addition to the limitation of \ours{} in the efficiency of trigger inversion mentioned in the previous section, we discuss other potential limitations of our work in this section.

Firstly, as illustrated in Section~\ref{sec:threat_model}, our defense mainly focuses on using third-party trained models. 
In particular, similar to existing baseline methods, we assume that defenders have a few local clean samples. 
Accordingly, our method is not feasible without clean samples. 
Besides, we need to train a model for the scenarios using third-party datasets before conducting trigger inversion and follow-up defenses, which is computation- and time-consuming. 
We will further explore how to conduct trigger inversion under few/zero-shot settings in our future works.

Secondly, similar to trigger inversion-based defenses in NLP~\cite{2022-Piccolo, 2022-Constrained-Optimization-with-Dynamic-Bound-scaling-for-Effective-NLP-Backdoor-Defense}, the trigger inversion process in our method also relies on a white-box setting. 
Accordingly, it does not apply to black-box scenarios in which the defenders can only access the final output of the backdoored model. 
We also note that in practical applications, it is often feasible to derive a white-box surrogate model from a black-box model using distillation techniques, as demonstrated in existing research~\cite{2024-Distilled-GPT-for-Code-Summarization, 2024-DistillSeq, 2024-Enhancing-Code-Generation-by-Distilling}. Once a white-box surrogate is obtained, \ours{} can be applied to mitigate backdoor vulnerabilities.
We will continue the exploration of designing black-box trigger inversion in our future works. 

\section{Conclusion}
\label{sec:conclustion}
In this paper, we propose \ours{}, a novel backdoor elimination technique for ensuring secure code understanding. By PL-specific trigger vocabulary generation and sample-specific trigger position identification, \ours{} reduces the search space for trigger optimization and minimizes the impact of non-backdoor perturbations, respectively. Our experiments show that \ours{} can effectively invert the trigger of the given backdoored NCM. Through trigger unlearning, \ours{} can reduce the average ASR of backdoored NCMs to a minimum of 0.24\% without impacting their performance on normal inputs.

\section{Data Availability}
\label{sec:data_availability}
Our source code and experimental data are available at~\cite{2025-EliBadCode}.

\section*{Acknowledgement}
We would like to thank anonymous reviewers for their insightful comments. This work is supported partially by the National Research Foundation, Singapore, and DSO National Laboratories under the AI Singapore Programme (AISG Award No: AISG2-GC-2023-008), the National Natural Science Foundation of China (61932012, 62372228, U24A20337), the Fundamental Research Funds for the Central Universities (14380029), the Open Project of State Key Laboratory for Novel Software Technology at Nanjing University (Grant No. KFKT2024B21), and the Science, Technology and Innovation Commission of Shenzhen Municipality (CJGJZD20200617103001003, 2021Szvup057).

\bibliographystyle{ACM-Reference-Format}
\bibliography{reference.bib}

\end{document}